# Localized Triggering of the Insulator-Metal Transition in $VO_2$ using a Single Carbon Nanotube

Stephanie M. Bohaichuk,[1] Miguel Muñoz Rojo,[1,2] Gregory Pitner,[1] Connor J. McClellan,[1] Feifei Lian,[1] Jason Li,[3] Jaewoo Jeong,[4] Mahesh G. Samant,[4] Stuart S. P. Parkin,[4] H.-S. Philip Wong,[1] Eric Pop[1,5,*]

[1]Dept. of Electrical Engineering, Stanford University, Stanford, CA 94305, USA

[2]Dept. of Thermal and Fluid Engineering, University of Twente, 7500 AE Enschede, Netherlands

[3]Asylum Research, Santa Barbara, CA 93117, USA

[4]IBM Almaden Research Center, San Jose, CA 95120, USA

[5]Dept. of Materials Science and Engineering, Stanford University, Stanford, CA 94305, USA

**ABSTRACT:** Vanadium dioxide ($VO_2$) has been widely studied for its rich physics and potential applications, undergoing a prominent insulator-metal transition (IMT) near room temperature. The transition mechanism remains highly debated, and little is known about the IMT at nanoscale dimensions. To shed light on this problem, here we use ~1 nm wide carbon nanotube (CNT) heaters to trigger the IMT in $VO_2$. Single metallic CNTs switch the adjacent $VO_2$ at less than half the voltage and power required by control devices without a CNT, with switching power as low as ~85 μW at 300 nm device lengths. We also obtain potential and temperature maps of devices during operation using Kelvin Probe Microscopy (KPM) and Scanning Thermal Microscopy (SThM). Comparing these with three-dimensional electrothermal simulations, we find that the local heating of the $VO_2$ by the CNT play a key role in the IMT. These results demonstrate the ability to trigger IMT in $VO_2$ using nanoscale heaters, and highlight the significance of thermal engineering to improve device behaviour.

*Contact: epop@stanford.edu

Materials with an abrupt insulator-metal transition (IMT) have garnered much interest, both as a study of the role of electron correlations in creating new electronic phases, and for their variety of potential applications in optics and electronics.[1,2] Vanadium dioxide ($VO_2$) has one of the most pronounced transitions among these, with a structural transition from monoclinic to rutile at ~340 K. This results in a drop in resistivity by up to five orders of magnitude, accompanied by significant changes in optical properties.[3,4] This transition can be induced electrically on a sub-nanosecond time scale by using current flow, and reverses once the stimulus is removed.[5] These properties have made $VO_2$ a candidate for threshold switches and selectors,[6-8] transistors,[9,10] oscillators,[11,12] and tunable metamaterials for optoelectronics.[13-16]

Integrating IMT materials with current semiconductor technology to build these applications will require knowledge of their behaviour at nanoscale dimensions. For example, electrical devices based on two-terminal switching of $VO_2$ are expected to offer faster,[5] lower voltage,[17-19] and lower energy[2] switching as they are reduced to smaller dimensions, similar to devices based on phase-change materials like $Ge_2Sb_2Te_5$.[20,21] Most two-terminal $VO_2$ devices studied to date have had dimensions ranging from ~20 nm to a few microns,[19] with IMT behaviour preserved in all cases. It remains to be seen if the IMT in electrical devices changes once even smaller dimensions are reached. Moreover, the nanoscale triggering mechanism of $VO_2$ is not completely understood, with some debate on the role of Joule heating[22] vs. electric field effects and carrier injection.[23,24] The distinction partly arises from the origin of the IMT (*e.g.* a Peierls structural transition triggered by heating and electron-phonon coupling *vs.* a Mott electronic transition based on carrier concentration), and will provide insight into the types of devices that can be designed.

In this work, we probe the mechanism of $VO_2$ switching at the nanoscale. To extend below the limits of lithography we use single-wall metallic carbon nanotube (CNT) heaters to trigger the $VO_2$ transition. Due to their ~1 nm diameter, such metallic CNTs are ideal candidates for probing a nanoscale phase change or IMT, as Joule heaters (capable of reaching ~600ºC in air, and ~2000ºC in vacuum) or electrodes.[25,26] By using the localized heating of a metallic CNT we are able to initiate the IMT at the nanoscale, at a lower power than relying on Joule heating in the $VO_2$ itself, which is promising for the development of applications requiring nanoscale $VO_2$ devices. We also use Kelvin Probe Microscopy (KPM) and Scanning Thermal Microscopy (SThM) to obtain high resolution spatial maps of the electric potential and temperature changes in our devices during operation, to understand their switching mechanism. We find good agreement between our experimental results and electrothermal simulations, confirming that Joule heating plays a major role in our devices, both with and without a CNT.

## RESULTS AND DISCUSSION

We fabricated nearly one thousand two-terminal $VO_2$ devices with and without CNTs on top. Aligned CNT arrays were grown on a separate quartz substrate, then transferred[27] onto thin films of single crystal $VO_2$ grown epitaxially[28] on $TiO_2$ (101), as illustrated in Figure 1a–d. The CNTs were coated with 100 nm of Au by electron-beam (e-beam) evaporation, then peeled off the quartz and transferred onto $VO_2$ using thermal release tape. The Au was wet-etched to leave only CNTs on $VO_2$ (see Methods for additional details). A scanning electron microscope (SEM) image of transferred CNTs on $VO_2$ is shown in Figure 1e. Excess CNTs were removed and the $VO_2$ was wet etched into stripes (Figure 1f, also see Methods). E-beam evaporated Pd contacts were added to make complete devices, as shown in Figure 1g,h.

We used electrical testing and atomic force microscopy (AFM) scans to select devices with single metallic CNTs for further study, and for comparison to control devices without a CNT. (The selection process, and comparisons with multiple-CNT or with semiconducting-CNT devices are described in the Supporting Information Section 2.) Figure 1g shows the schematic of a $VO_2$ device with a single CNT heater, both extending underneath Pd contacts. A series resistor $R_S$ is used as a current compliance to protect devices from overheating failure in the metallic state, and to reduce current overshoot from external capacitance (*e.g.* cables and probe arm). The $R_S$ value (20 – 200 kΩ) is chosen to be a small fraction of the insulating $VO_2$ resistance, but higher than the metallic $VO_2$ resistance, as detailed in the Supporting Information Section 2. Figure 1h shows an optical image of a shorter device, fabricated by adding Pd contact extensions. The measured resistance of a $VO_2$ device without a CNT is shown in Figure 1i as a function of stage temperature, displaying the transitions at 328 K and 321 K for heating ($T_{IMT}$) and cooling ($T_{MIT}$) respectively, and a change in resistance over three orders of magnitude.

Figure 2a compares typical measured voltage-controlled *I-V* characteristics of $VO_2$ devices with and without a CNT ($L$ = 6 μm) at room temperature ($T_0$ = 296 K). Electrical switching is repeatable and independent of bias polarity, with similar behaviour consistently observed across hundreds of devices. The non-CNT device behaves linearly as a resistor, until significant Joule heating begins to occur. As the $VO_2$ temperature increases, it becomes more conductive and the *I-V* curve is increasingly superlinear. Once the transition occurs at a critical voltage $V_{IMT}$, most of the applied bias is dropped across $R_S$, causing a snapback in device voltage. In the metallic state $R_S$ dominates over the $VO_2$ resistance, limiting the maximum current, power, and on/off ratio observed. Because $R_S$ is used to limit heating, if devices were operated at much shorter time scales, then $R_S$ could likely be reduced, recovering more of the intrinsic on/off ratio of the $VO_2$.

$R_S$ can also be used to control the resistance and volume of the $VO_2$ that is metallic (Supporting Information Figure S8).

There are significant differences in the *I-V* characteristics when a single metallic CNT is present. Prior to the IMT there is higher current and a sublinear behaviour typical of current saturation due to self-heating in the CNT.[29] The IMT of the $VO_2$ occurs at much lower power, because the (hot) CNT is able to switch a highly localized $VO_2$ region at significantly lower voltage compared to Joule heating through the entire $VO_2$. Once an initial region of $VO_2$ below the CNT has switched, the increased current from the metallic $VO_2$ becomes self-sustaining and the metallic region can expand, leading to a large and abrupt increase in current. The metal-insulator transition (MIT) that occurs when the voltage is ramped back down is unaffected by the CNT, which no longer carries the majority of the current once the voltage snapback occurs. Hysteresis is observed in both types of devices because $T_{IMT} \neq T_{MIT}$ and because at a given voltage, metallic $VO_2$ will cause more heating ($\propto V^2/R$) than insulating $VO_2$. Due to the reduction in $V_{IMT}$, devices with a CNT have significantly smaller total hysteresis window. As expected, both types of devices show a decrease in $V_{IMT}$ with rising ambient temperature (Supporting Information Section 2). At all temperatures measured, devices with a CNT display lower switching voltage and power compared to control devices without a CNT.

These differences in *I-V* characteristics of $VO_2$ devices with and without a CNT are also seen at shorter length scales (*i.e*. Pd contact separation), shown in Figure 2b. Switching is consistently triggered by the ~1 nm wide CNT at all length scales, shown in Figured 2c–d. The presence of a CNT halves the required switching voltage and power in all devices measured, including our shortest 300 nm lengths. Figure 2c shows that switching voltage scales linearly with length for both devices types. (Width scaling of our $VO_2$ devices is displayed in Supporting Information Figures S4 and S15.) Shorter devices have lower resistance and higher Joule heating at a given voltage, thus requiring a lower voltage and power for switching. An effective switching field can be extracted from the slope of the linear fits in Figure 2c, giving $3.5 \pm 0.2$ V/µm with a CNT and $7.6 \pm 0.2$ V/µm without, though this does not necessarily indicate a field-switching mechanism. If switching were triggered by field effects such as carrier injection, then the field extracted would be a description of the $VO_2$ quality and the efficiency of the switching mechanism. For a Joule heating mechanism, the field would be determined by the electrical and thermal properties of the materials that set the maximum device temperature (including ambient temperature, uniformity of heating, thermal conductivities and thermal boundary resistances, resistivities, *etc*.).

The vertical axis intercept in Figure 2c (2.0 ± 0.8 V with a CNT, and 6.8 ± 1.0 V without) characterizes the voltage drop (contact resistance) and heat loss at the contacts.[19] This intercept depends on the contact material[17] and its temperature-dependent contact resistivity.[30,31] The large difference between the intercept of devices with and without a CNT is likely due to a lower contact resistance to the CNT than the $VO_2$. An estimate of the contact resistance at switching can be found using the intercept and typical switching currents (Supporting Information Figure S3), giving 32 ± 15 kΩ and 123 ± 48 kΩ for devices with and without a CNT, respectively. This is consistent with other estimates (Supporting Information Section 2),[29, 32] but due to non-uniform current flow a full interpretation of these values is difficult. The ratio between the voltage drop at the contacts and the effective switching field yields a characteristic length below which the switching voltage could be contact-dominated, ~0.6 µm and ~0.9 µm in our devices with and without a CNT respectively. Although the power reduction observed in devices with a CNT comes partly from localized switching reducing the required field, a significant part comes from the difference in contacts, which could limit the switching voltage and temperature in nanoscale thin film devices.

Given that the switching current is similar between devices with and without CNTs for the device dimensions used (Supporting Information Figure S3), Figure 2d shows that there is a reduction in power at all length scales by using a CNT. Normalization by $VO_2$ width is appropriate for devices without a CNT, but adds spread in the $P_{\mathrm{IMT}}$ of devices with a CNT, whose switching does not depend on the $VO_2$ width. Power scales linearly with device length, and our shortest devices with and without a CNT have switching powers of 85 µW and 260 µW respectively, among the lowest reported at similar $\Delta T = T_{\mathrm{IMT}} - T_0$.[6, 19] It is expected that further reducing our device length and width would reduce switching power. These results demonstrate the feasibility of $VO_2$ switching down to the nanoscale, with its IMT behaviour preserved, and that there are power benefits to doing so.

The debate regarding thermal and non-thermal IMT effects prompts us to examine whether our electrical results can be explained solely by Joule heating. To gain insight into the switching mechanism of devices with a CNT, we utilize KPM and SThM scanning probe techniques. KPM is a non-contact scanning probe technique that detects changes in the surface potential across a sample.[33] We use KPM to study the potential in biased $VO_2$ devices with and without a CNT. On the other hand, SThM is a contact-mode scanning probe technique that uses a thermo-resistive probe sensitive to temperature changes on the surface of a sample.[34,35] We use SThM to study the heating profile of biased devices in order to identify the thermal contribution of the CNT to the IMT of the $VO_2$.

Figure 3a shows a topographic scan of a $VO_2$ device without a CNT, and Figures 3b–d show KPM results for that device, with applied voltages $V_S$ as labelled. Scans are centered on the $VO_2$ channel with the $TiO_2$ substrate revealed along the left and right edges. The small spots are carbon-based residue from processing. The contacts are just outside the scan with the grounded electrode at the top and the positive electrode at the bottom, connected to $R_S$ = 200 kΩ. Figure 3b shows a KPM scan with no bias across the device. The $VO_2$ appears uniform, with a slight contrast against the process residues and $TiO_2$. Figures 3c and 3d show KPM scans taken with a constant voltage $V_S$ applied, (c) in the insulating state and (d) once the $VO_2$ has electrically switched to the metallic state. In biased devices, there is a linear decrease in potential from the positive electrode to the grounded electrode (see Supporting Information Section 3). The scans have been processed with a first order line flattening operation to remove this, highlighting local differences in surface potential across the device width. In both states, the device has for the most part a nearly uniform potential drop across it with no strongly localized fields. In the metallic state some slight variation exists across the width of the device from local differences in temperature and conductivity.

Figure 3e shows a topographic scan of a $VO_2$ device with a CNT, and Figures 3f–h show KPM results for the device, with potential variation across the width that differs significantly from the non-CNT device. The orientation of the electrodes relative to the images is the same, with the $VO_2$ edges just outside the scan on the left/right. These images were processed in the same way. Although the raw potential drop is linear, flattening reveals a small local concentration of the surface potential around the CNT in the insulating state (g). Once the device switches to the metallic state (h) and voltage snapback occurs, the flattened potential appears much more uniform across the device, with the CNT having little effect anymore.

The contrast in the potential across the devices with and without a CNT indicates that the CNT has a large impact on the $VO_2$ switching. To test whether this can be attributed to thermal effects, we perform SThM on a similar $VO_2$ device with a CNT, which has a spatial resolution of <100 nm. Figures 4a and 4b show topographic scans of this device before and after capping with a 50 nm layer of poly(methyl methacrylate) (PMMA), with the CNT no longer visible after capping. This PMMA layer is needed for electrical insulation between the SThM probe and sample surface. The contacts are at the top and bottom of the image, and the device is held at a constant voltage, $V_S$ = 17 V, with $R_S$ = 200 kΩ. The SThM results in Figure 4c prior to switching confirm that there is significant localized heating around the CNT.

To quantify the local temperature rise in the $VO_2$ induced by the CNT, we perform three-dimensional (3D) finite element simulations which self-consistently consider electrical, thermal, and Joule heating effects. The electrical conductivity of both the CNT and the $VO_2$ are described as a function of temperature (Supporting Information Section 4). We also include electrical contact resistances and thermal boundary resistances, which cannot be neglected. The simulated device has the same dimensions as the real device scanned by SThM, capped by 50 nm of PMMA with a CNT at the device center. Simulating the device at the same bias as the SThM scan, we see in Figure 4d a similar temperature profile on the PMMA surface compared to the real device. Figure 4e shows that we can reproduce the experimental *I-V* curve, with the simulated device having $V_{IMT}$ ~ 20 V. Figure 4f shows the simulated temperature profiles on the $VO_2$ and PMMA surface at the SThM bias point, in the center of the device perpendicular to the CNT. The peak temperature in the $VO_2$ directly underneath the CNT is higher and the temperature rise much more laterally confined than observed on the PMMA surface. Only a few-nanometer wide $VO_2$ region under the CNT will reach $T_{IMT}$ and trigger the transition, compared to nearly the entire device width when a CNT is not present (Supporting Information Figure S14a). The CNT itself is much hotter, reaching a temperature of ~400 K, but the thermal boundary resistance (between CNT and $VO_2$) and small contact area limit heat flow from the CNT to the $VO_2$.

When the series resistance $R_S$ (= 200 kΩ) is added to the model to limit positive feedback, the simulation can also reproduce switching to the metallic state, as shown in Figure 5a for a device with a CNT. These simulations show that the metallic $VO_2$ forms a narrow conducting "filament," ~10 nm wide, just beneath the CNT. Switching is always triggered beneath the CNT regardless of its location in the $VO_2$ channel, meaning that using a localized heater can provide a means of control over switching location. Full *I-V* curves for both devices with (Figure 5b) and without (Figure 5c) a CNT can be simulated by sweeping the voltage, where the downwards sweep uses the cooling branch of the $R(T)$ curve (Figure 1b) to model hysteresis. Both curves reproduce experimental *I-V* behaviour remarkably well, including the differences in switching voltage and hysteresis between the two types of devices, using only Joule heating in the model and no other field effects. Thus combined, our simulations, KPM, SThM, and electrical results suggest that Joule heating is a valid explanation for the mechanism of switching in our devices. Although a thermally-assisted field mechanism cannot be excluded using our data (for example, heating can increase carrier concentration to trigger a Mott transition or reduce the energy barrier for field-induced switching), Joule heating plays a key role in switching devices even in the narrow $VO_2$ region activated by the hot CNT.

## CONCLUSIONS

In summary, we have shown that the IMT switching of $VO_2$ can be triggered by nanoscale heaters made of individual metallic carbon nanotubes. Two-terminal $VO_2$ devices with CNTs exhibit switching at less than half the voltage and power of traditional $VO_2$ devices without a CNT, at all length scales. Using a combination of scanning probe techniques and finite element simulations we studied the origin and scale of the IMT in such devices with and without CNT heaters. Our results are consistent with a Joule heating mechanism in which the CNT locally heats the $VO_2$ and triggers IMT in narrow region. These results highlight the importance of thermally engineering devices for low-power switching, by using confined heating in small volumes, and are also applicable to a wide variety of thermally-activated phase-change and resistive switching devices.

## METHODS

Thin films of single crystalline $VO_2$ are epitaxially grown on $TiO_2$ (101) substrates using pulsed laser deposition (PLD), with a nominal thickness of 9 nm.[28] Separately, we grow aligned CNTs with an average diameter of 1.2 nm by chemical vapor deposition (CVD) on ST-cut quartz, then transfer them onto the $VO_2$.[27] The CNTs on quartz were coated with 100 nm of Au by electron beam (e-beam) evaporation, onto which thermal release tape was pressed (Semiconductor Equipment Corp 1398MS, with adhesion 2.5 N / 20 mm and release temperature 120ºC). The CNT/Au/tape stack was peeled off the quartz and then pressed onto the $VO_2$. The tape was released on a hot plate at 130ºC, leaving behind the Au-coated CNTs on the $VO_2$ surface. An $O_2$ plasma clean (20 sccm, 25 mTorr, 55 W, 3 min) followed by an Ar plasma clean (15 sccm, 12.5 mTorr, 100 W, 3 min) were done to remove tape residue on the Au, with the $VO_2$ protected from damage by the Au film. The remaining Au was removed using a KI wet etch, leaving behind aligned CNTs on the $VO_2$. Some carbon-based residue is left after the transfer process (Figure 1e).

The $VO_2$ and CNTs were patterned into stripes of width $W$ = 3 to 9 μm using a photoresist etch mask. CNTs outside the $VO_2$ stripes were removed using a light $O_2$ plasma (20 sccm, 150 mTorr, 30 W, 1 min), then the $VO_2$ was wet etched for 30 s using a 25% nitric acid solution. Two-terminal devices were fabricated with 50 nm thick Pd contact pads (with no Ti sticking layer) with dimensions 300 μm × 250 μm *via* e-beam evaporation and lift-off, with spacing (device lengths) ranging from $L$ = 3 to 10 μm (Figure 1g). Shorter devices with $L$ = 300 nm to 2 μm were made by adding small extensions of 50 nm thick Pd to the existing pads using e-beam lithography. Presence of metallic CNT(s) in devices was verified electrically and the number of CNTs confirmed by atomic force microscopy (AFM). The $VO_2$ film thickness after all processing and etching steps is ~5 nm measured by AFM.

Electrical measurements are performed in a micromanipulator probe station from Janis Research under vacuum (<$10^{-4}$ Torr) with a Keithley 4200-SCS parameter analyzer applying a voltage $V_S$, all at room temperature ($T_0$ = 296 K) unless otherwise stated. A series resistance $R_S$ = 20 kΩ and 100 kΩ is used for short (< 2 μm long) devices with and without CNTs respectively, and 200 kΩ is used for all other devices. KPM is done on an Asylum Research system with a high voltage module, while the device is biased at a constant voltage. Devices are coated with 50 nm thick 2% 495K PMMA in anisole to carry out passive-mode SThM measurements. PMMA is used rather than an oxide capping layer, because oxide deposition can reduce the stability of the CNT and $VO_2$. The SThM tip (Pd on SiN, model PR-EX-GLA-5 from Anasys®) is a thermo-resistive element sensitive to electrical discharges, so this capping is necessary in order to electrically isolate the tip from the device while it is biased. SThM is done in passive mode, with a 0.5 V set point and a 0.5 V tip bias.

## ASSOCIATED CONTENT

### Supporting Information

Additional details of CNT growth and VO2 characterization, additional electrical measurements, discussion of device scaling, discussion of semiconducting and multi-CNT devices, discussion on the use of the series resistor, discussion of contact resistance, additional KPM images, all COMSOL simulation details.

## ACKNOWLEDGMENTS

The authors gratefully acknowledge Eilam Yalon and Suhas Kumar for commenting on the manuscript. This work was supported in part by the Stanford SystemX Alliance and by the National Science Foundation (NSF). Work was performed in part at the Stanford Nanofabrication Facility and the Stanford Nano Shared Facilities which receive funding from the NSF as part of the National Nanotechnology Coordinated Infrastructure Award ECCS-1542152. S.B. acknowledges support from the Stanford Graduate Fellowship (SGF) program and the NSERC Postgraduate Scholarship program.

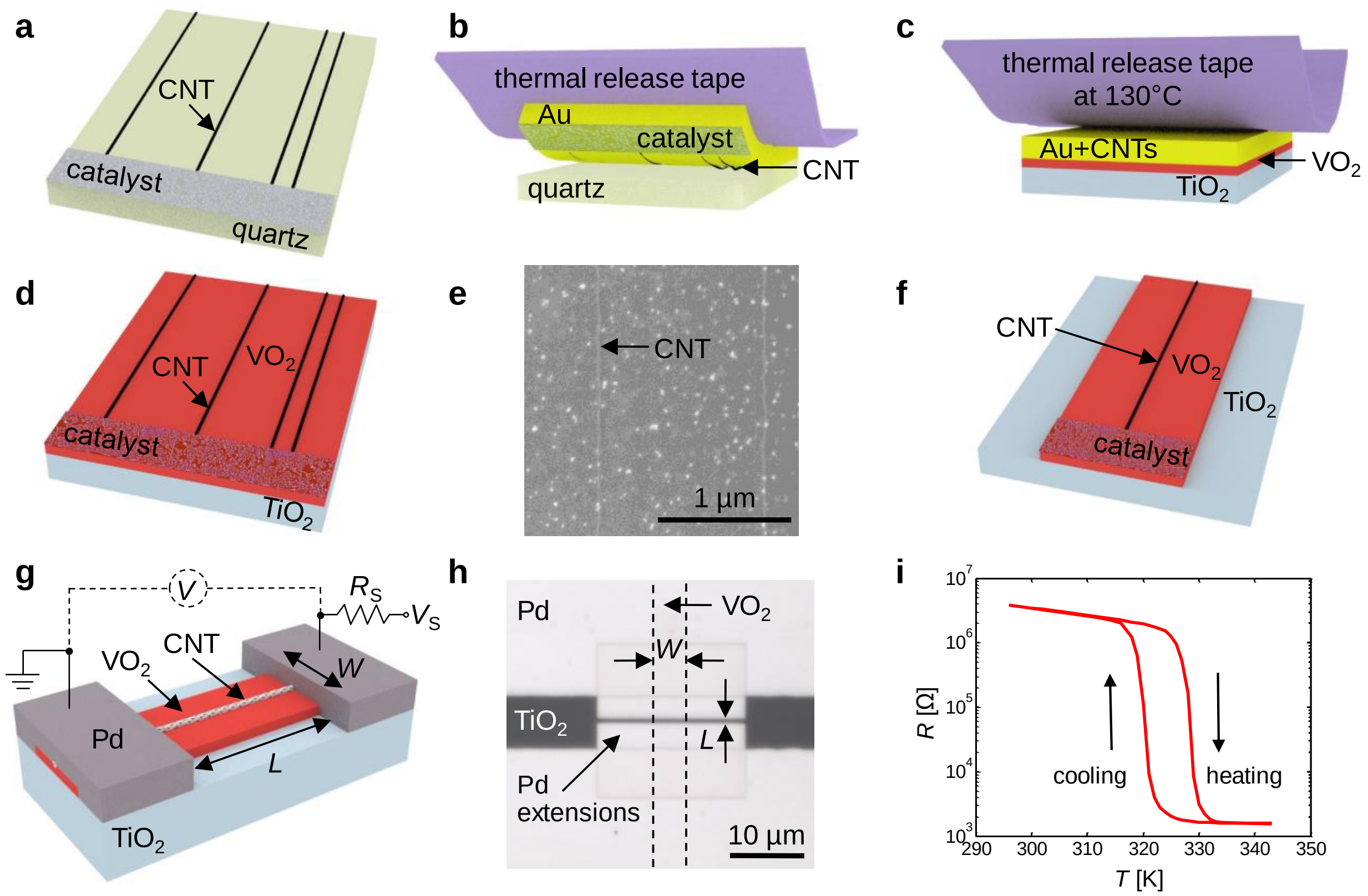

**Figure 1.** Fabrication Process (layers are not to scale). (a) Aligned carbon nanotubes (CNTs) are grown on ST-cut quartz *via* CVD, with Fe catalyst particles.[27] (b) Au is e-beam evaporated onto the CNTs, then peeled off the quartz using thermal release tape. (c) The Au-coated CNTs are pressed onto $VO_2$ grown epitaxially by PLD on $TiO_2$.[28] The thermal release tape is then removed by heating to 130°C. (d) After a plasma clean to remove tape residue, the Au is wet-etched, leaving behind CNTs on $VO_2$. (e) Scanning electron microscope (SEM) image of CNTs on $VO_2$. The small dots are residue left by the transfer process. (f) Excess CNTs outside planned $VO_2$ stripes are removed with an $O_2$ plasma then the $VO_2$ is wet etched in diluted nitric acid. (g) Schematic of a fabricated $VO_2$ device with CNT heater and measurement setup, after e-beam evaporation and lift-off of Pd contacts. *W* is the width of the patterned $VO_2$ region and *L* is the separation between Pd contacts, *i.e.* the $VO_2$ channel length. The thickness of the $VO_2$ is ~5 nm in finished devices after all processing steps. Similar control devices were fabricated without CNT heaters. (h) For shorter devices, additional Pd contact extensions are added. Optical image of a short ($L$ = 520 nm, $W$ = 3.9 μm) $VO_2$ device. (i) Measured resistance of a $VO_2$ device without a CNT heater as a function of stage temperature. ($L$ = 5 μm, $W$ = 7 μm)

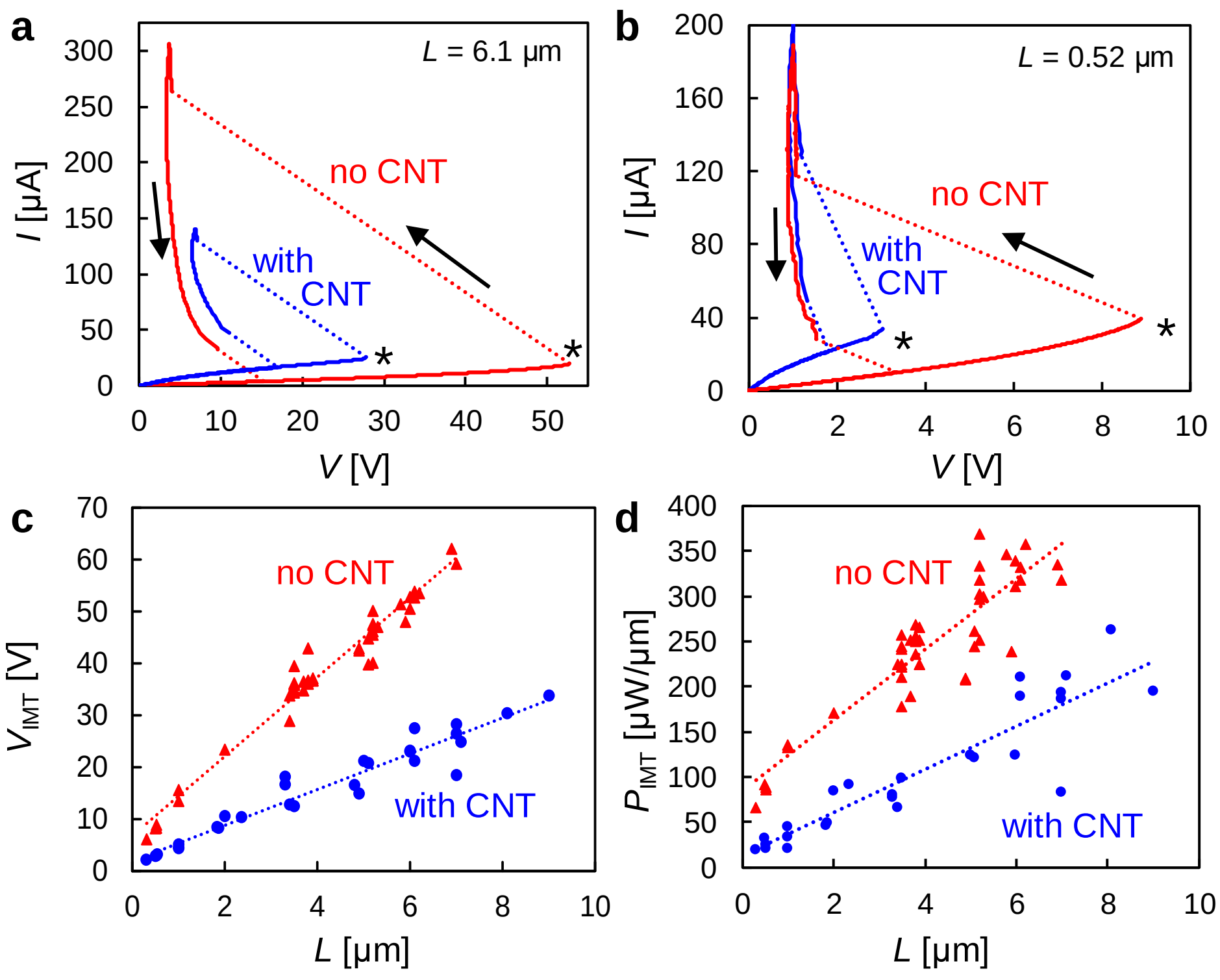

**Figure 2.** Electrical switching of devices with and without a CNT. (a) Typical switching of "long" devices with and without a CNT ($L$ = 6.1 μm, $W$ = 3.2 μm) using DC voltage control. (b) Typical switching of "short" devices with and without a CNT ($L$ = 520 nm, $W$ = 4 μm) using DC voltage control. The short devices and the devices with a CNT heater have much lower switching voltages, $V_{IMT}$, labeled with * on the figures. Arrows show voltage sweep directions, and dashed lines indicate snapbacks. (c) Measured $V_{IMT}$ as a function of length for devices with (blue circles) and without (red triangles) a CNT. Dotted lines represent a linear fit. (d) Switching power, $P_{IMT}$, normalized by $VO_2$ width, for devices with (blue circles) and without (red triangles) a CNT. The dotted lines represent a linear fit. Adding a CNT approximately halves switching power at all length scales.

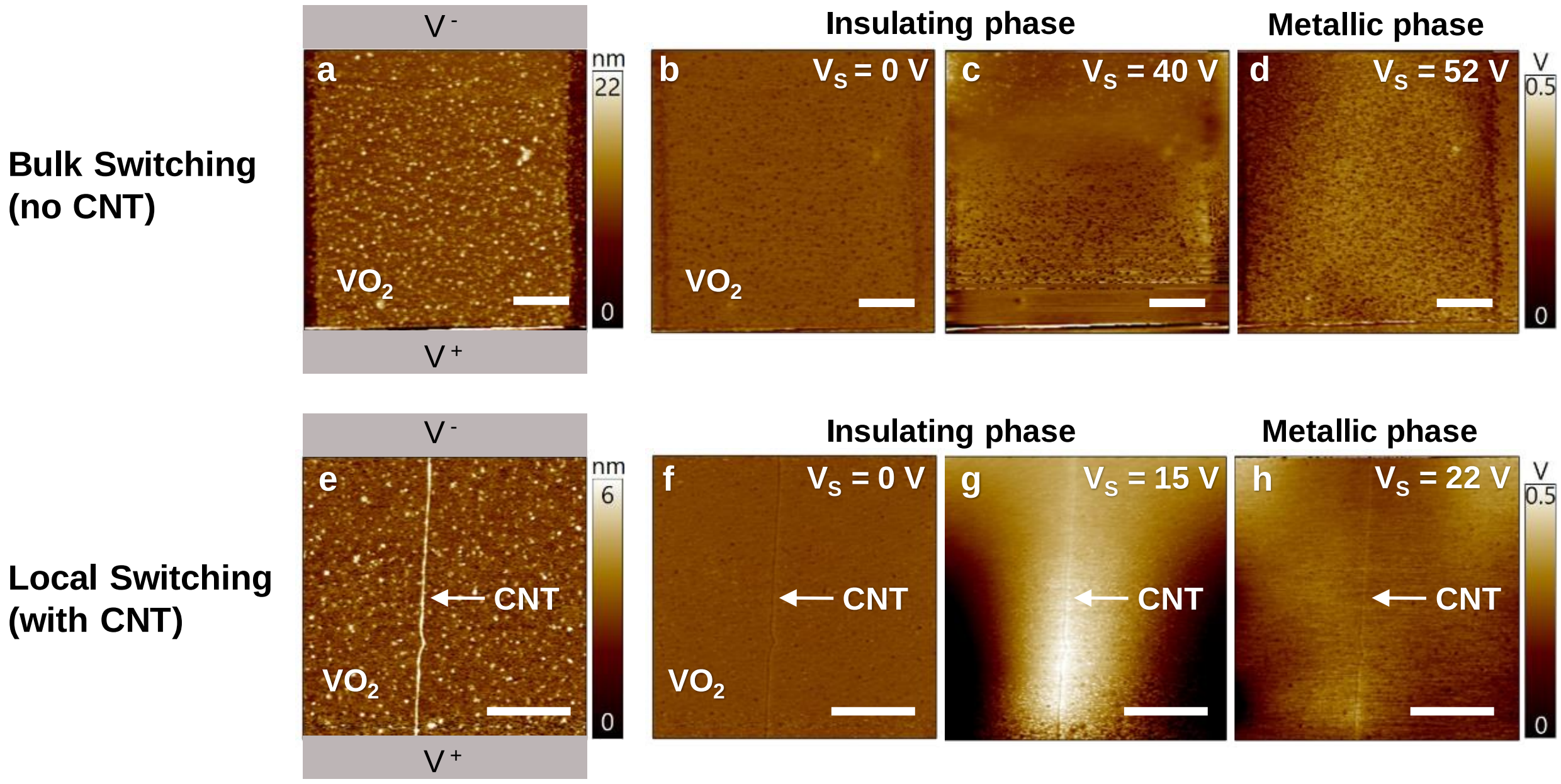

**Figure 3.** Topography and Kelvin Probe Microscopy (KPM) of $VO_2$ devices with and without CNT. (a) Topography of $VO_2$ control device without a CNT. (b – d) Flattened KPM images of the same device with increasing bias. The Pd electrodes are outside the top and bottom margin of the device images, biased as marked. (e) Topography of $VO_2$ device with a single metallic CNT, indicated by the arrow. (f – h) Flattened KPM images of the same device with increasing bias, revealing localized switching caused by the CNT. All scale bars are 1 μm.

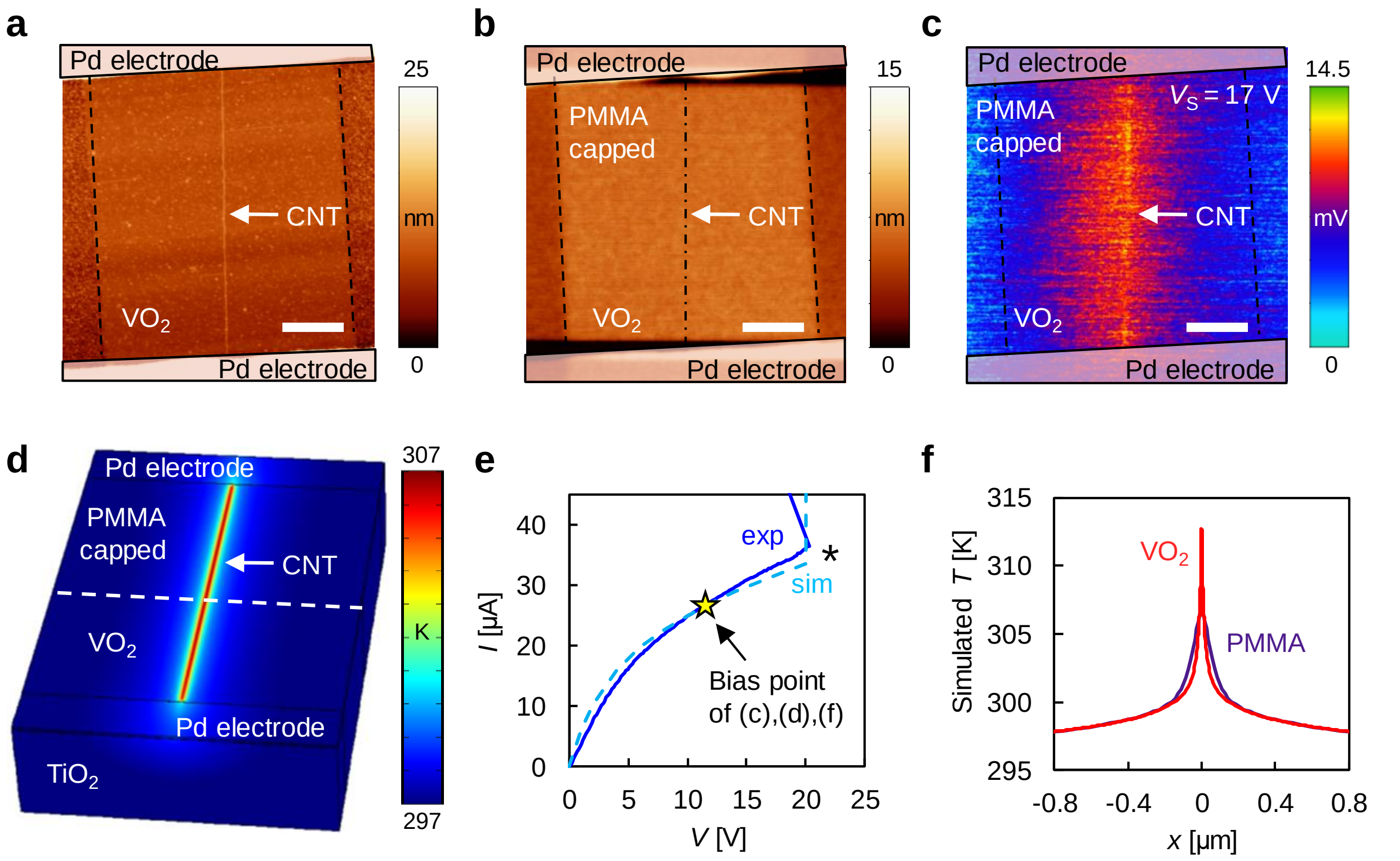

**Figure 4.** (a) Topography of an unbiased device with a single metallic CNT. (b) Topography of the same device covered in 50 nm of PMMA, with the CNT no longer visible. (c) Scanning thermal microscopy of the device under bias, prior to the metallic transition which occurs at $V_S$ = 27 V ($V_{IMT}$ = 20.3 V). (d) Simulated surface temperature of the device on top of the PMMA, at the same bias voltage. (e) *I*-*V* characteristics of the device (solid blue) compared to the model (dashed light blue), with switching marked by a *. (f) Simulated temperature profile across the $VO_2$ (red) and on the PMMA surface (purple) perpendicular to the CNT along the dashed white line in (d). Scale bars are 1.5 μm.

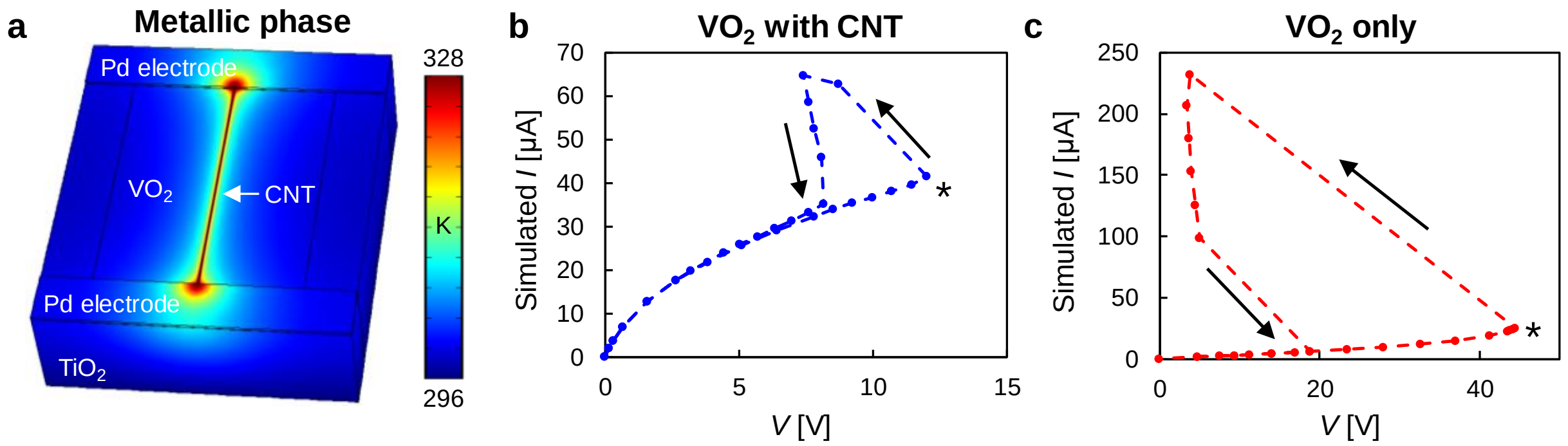

**Figure 5.** (a) Simulated temperature on the surface of a device with a CNT ($L$ = 5 µm, $W$ = 4 µm) after the metallic transition, with $R$s = 200 kΩ. (b) Simulated $I$-$V$ curve of a device with a CNT using voltage control, including hysteresis. (c) Simulated $I$-$V$ curve of a device without a CNT using voltage control, including hysteresis. The switching voltage (*) is much higher without the CNT to act as a heater.

# Localized Triggering of the Insulator-Metal Transition in $VO_2$ using a Single Carbon Nanotube

Stephanie M. Bohaichuk,[1] Miguel Muñoz Rojo,[1,2] Gregory Pitner,[1] Connor J. McClellan,[1] Feifei Lian,[1] Jason Li,[3] Jaewoo Jeong,[4] Mahesh G. Samant,[4] Stuart S. P. Parkin,[4] H.-S. Philip Wong,[1] Eric Pop[1,5,*]

[1]Dept. of Electrical Engineering, Stanford University, Stanford, CA 94305, USA

[2]Dept. of Thermal and Fluid Engineering, University of Twente, 7500 AE Enschede, Netherlands

[3]Asylum Research, Santa Barbara, CA 93117, USA

[4]IBM Almaden Research Center, San Jose, CA 95120, USA

[5]Dept. of Materials Science and Engineering, Stanford University, Stanford, CA 94305, USA

[*]Contact: epop@stanford.edu

## Supplementary Information Content:

## 1. $VO_2$ Characterization and CNT Growth

$VO_2$ thin films 9 nm thick were deposited using pulsed laser deposition on 1 $cm^2$ single crystal $TiO_2$ (101) pieces.[1] As-deposited $VO_2$ films were smooth with an rms roughness of ~1 Å. X-ray photoelectron spectroscopy (XPS) indicated the presence of a very thin surface oxidation, likely $V_2O_5$ (Figure S1).

Separately, horizontally aligned single-wall carbon nanotubes (CNTs) were grown via chemical vapor deposition (CVD) on ST-cut quartz using an ethanol source, which resulted in CNTs with an average diameter of 1.2 nm.[2] The CNTs grew perpendicular to patterned stripes of 1.3 Å thick Fe catalyst nanoparticles, resulting in a mixture of semiconducting and metallic CNTs with a random spacing between them, on average 0.3 CNT/µm. This density was chosen so that fabricating 3 to 9 µm wide $VO_2$ channels would result in a mixture of devices with no CNT, a single CNT, and multiple CNTs.

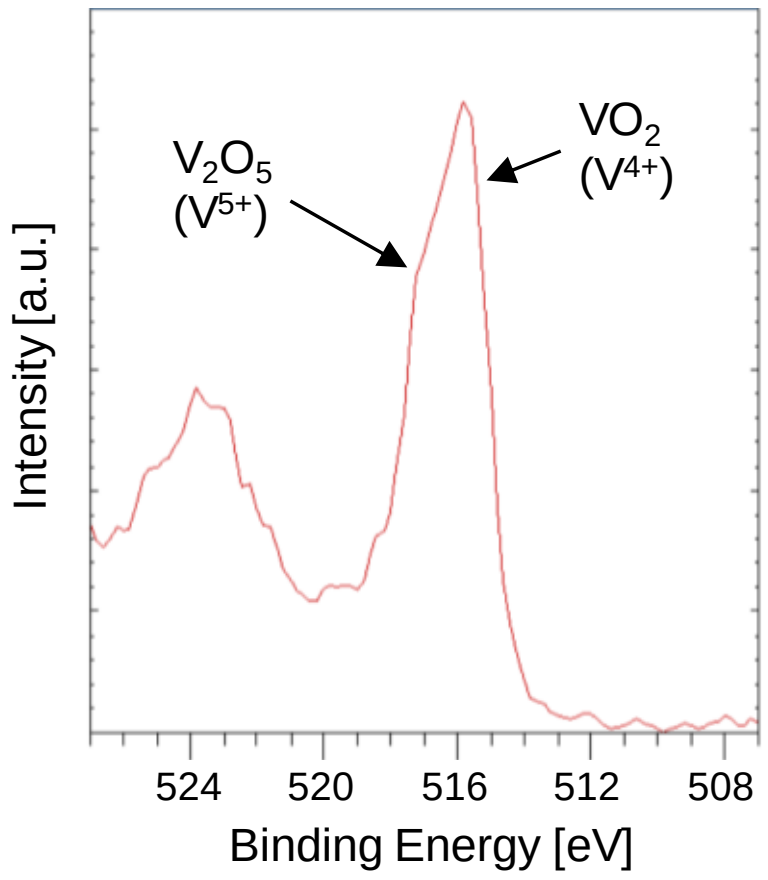

**Figure S1.** X-ray photoelectron spectroscopy (XPS) of the $VO_2$ film prior to processing, with V2p3 peaks indicating a $VO_2$ composition with a slight surface oxidation.

## 2. Electrical Measurements of Devices

### 2a. Additional Electrical Characterization

Measurements shown were all performed under vacuum ($< 10^{-4}$ Torr) for consistency, removing potential effects of atmosphere or water vapour on the CNTs. However, measurements can safely be done in air and yield nearly identical results. Figure S2 shows that devices have the same switching voltage and *I-V* characteristics regardless of bias polarity. Conduction is linear at low bias in either direction, indicating ohmic contacts.

Figure S3a shows that switching currents ($I_{IMT}$) are similar between devices with and without a CNT for the range of widths studied. For devices without a CNT, switching depends on heating of the $VO_2$ (which occurs through the entire width) and $I_{IMT}$ is thus higher in thicker or wider devices. This accounts for the spread of $I_{IMT}$ seen in devices without a CNT in Figure S3a. In devices *with* a CNT we expect the switching current to be independent of $VO_2$ width, because the narrow CNT provides the same heating regardless of the width of the $VO_2$ stripe underneath. Figure S3b shows switching current normalized by $VO_2$ width, which is appropriate for devices without a CNT but adds spread in the $I_{IMT}$ of devices with a CNT.

A slight increase in device switching voltage ($V_{IMT}$) is observed in narrow devices without a CNT, as shown in Figure S4a (accompanying simulations are in Section 4b). After burn-in, which is larger for narrower devices (Section 2e), $V_{IMT}$ has a much weaker width dependence across the range of device widths measured (3 to 9 µm). As the $VO_2$ width is reduced the switching current and power decrease slightly, shown in Figure S4b and S4c, respectively. In contrast, devices with a CNT show no dependence of $V_{IMT}$ on $VO_2$ width, and depend solely on the CNT quality and the efficiency of heat transfer to the $VO_2$.

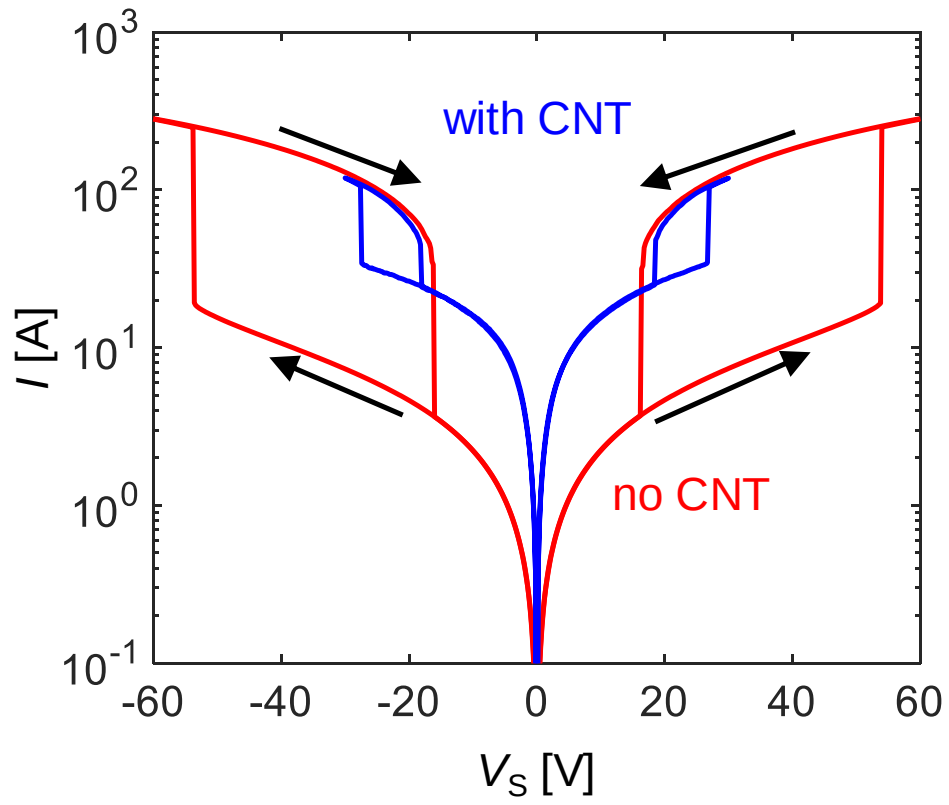

**Figure S2.** Forward and reverse DC voltage-controlled switching of devices with ($L$ = 6 µm, $W$ = 2.9 µm) and without a CNT ($L$ = 6 µm, $W$ = 5.1 µm), showing switching is independent of polarity.

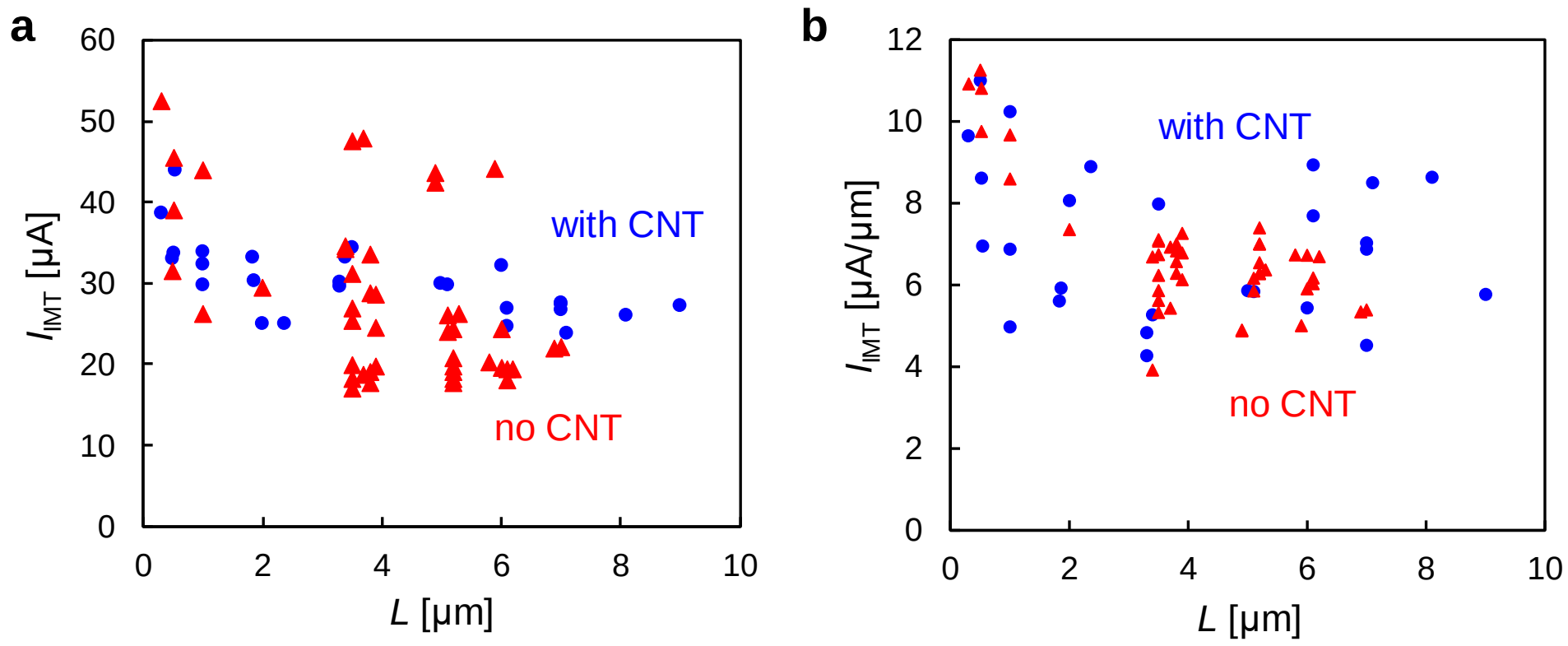

**Figure S3.** (a) Total current just prior to switching ($I_{IMT}$) for devices with (blue circles) and without (red triangles) a CNT, for a variety of widths ($W$ = 3 to 9 µm). (b) $I_{IMT}$ normalized by $VO_2$ width $W$, for devices with (blue circles) and without (red triangles) a CNT.

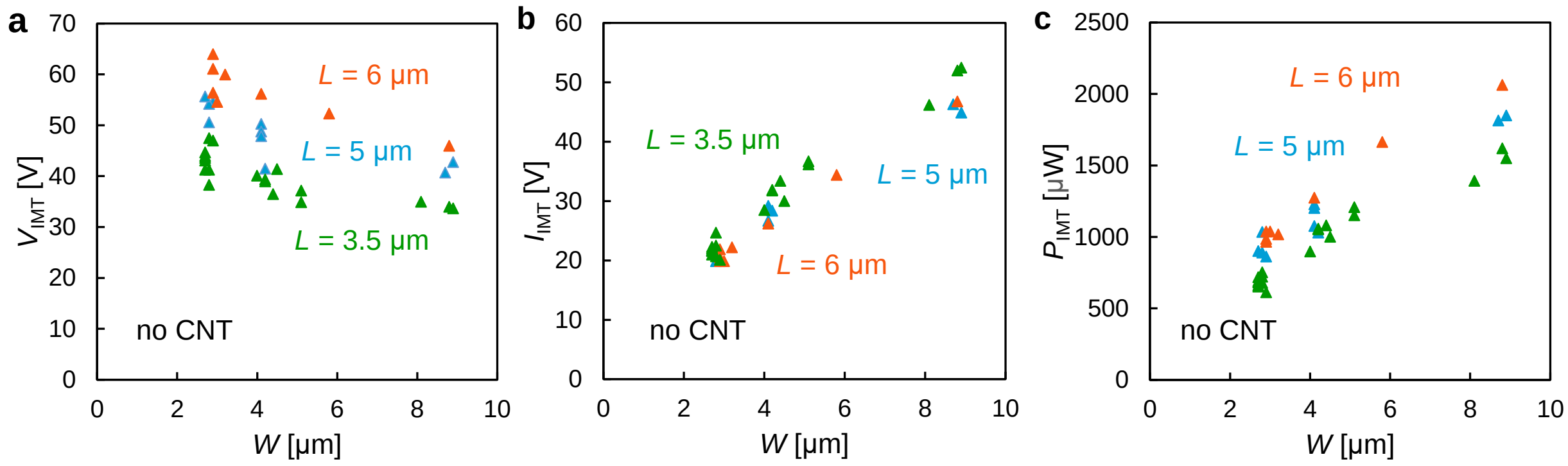

**Figure S4.** (a) Switching voltage $V_{IMT}$ for devices without a CNT is higher in narrower devices during the first switching event. After burn-in (Section 2e), $V_{IMT}$ shows a much weaker trend with width. (b) Switching current $I_{IMT}$ for devices without a CNT scales linearly with width. (c) Switching power of devices without a CNT is lower in narrower devices.

## 2b. Semiconducting and Multi-CNT Devices

The CNT growth conditions on quartz generate a mixture of semiconducting and metallic CNTs. In addition, because the spacing between parallel CNTs is uneven, our process ultimately yields a range of $VO_2$ devices with semiconducting CNT(s), with metallic CNT(s), with a mixture of both, or with no CNTs. The presence of metallic or "weakly metallic" CNT(s) was determined by the current carried in two-terminal electrical measurements, and the number of CNTs was confirmed by atomic force microscopy (AFM). A metallic CNT carries considerably more current and will show evidence of sublinear *I*-*V* behaviour (typically beyond ~ 3 V when $L$ ~ a few microns) and current saturation at higher voltages (typically near 10 to 25 μA).[3] Weakly metallic CNTs (damaged or poorly contacted metallic CNTs, or CNTs with a small band gap) are less conductive and often without clear saturation. Semiconducting CNTs carry much less current (typically ≪ 1 μA ungated), and their presence was verified through AFM. Examples of metallic and "weakly metallic" CNT *I*-*V* characteristics are shown in Figure S5a, with CNT devices fabricated in a similar manner but on $TiO_2$ (001) with no $VO_2$ film present (the substrate is insulating). Although the focus of this study was on devices with single metallic CNTs, the switching behaviours of several other types of devices were also measured. Figure S5b shows the switching voltages ($V_{IMT}$) of 112 devices, including devices with CNT(s) (filled circles) and without (open triangles).

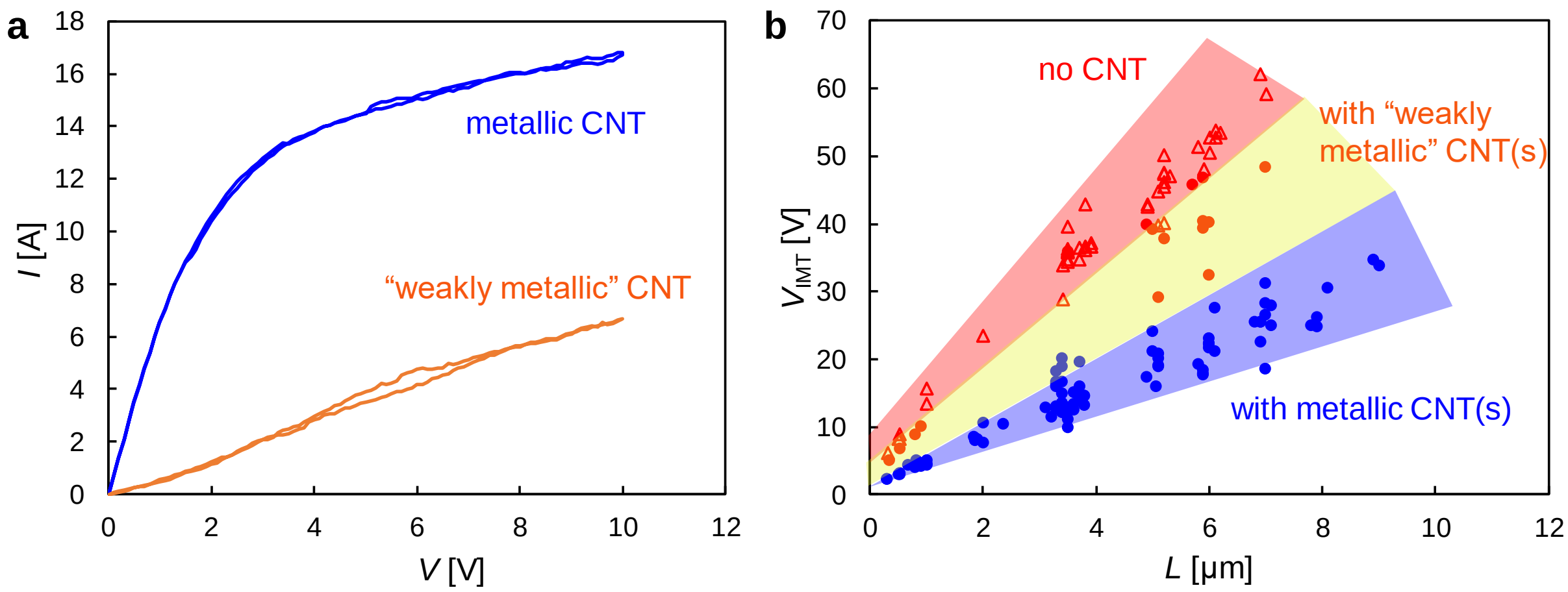

**Figure S5.** (a) Example *I*-*V* curves of devices ($L$ ~ 4 μm) with metallic and "weakly metallic" CNTs on $TiO_2$ (no $VO_2$ film). (b) Switching voltage $V_{IMT}$ as a function of length (contact separation) for devices with no CNT (open triangles) and at least one CNT (filled circles). Devices with *I*-*V* curves that indicate the presence of metallic CNT(s) are in blue, and those without are in red. Those that show a "weakly metallic" CNT presence in their *I*-*V* characteristics (orange) have a switching voltage in between the two groups.

Devices with a semiconducting CNT (red circles) have the same switching voltage as devices without any CNTs (red triangles). These devices are electrically indistinguishable from each other, because the ungated semiconducting CNTs do not carry sufficiently high currents to induce heating in the $VO_2$.

Devices with multiple metallic CNTs switch at the same voltage as those with single metallic CNTs (both blue circles). This suggests that the transition is triggered by the "best" (i.e. most conductive) metallic CNT, which heats up the most. After the $VO_2$ voltage snapback (due to the series resistor $R_S$) the CNTs no longer carry enough current to trigger further IMT, so any benefit to having multiple CNTs is lost. Note that all our devices have CNTs spaced far apart (~0.3 µm on average), and thus each CNT acts independently from one another. If the metallic CNTs are spaced much closer, $V_{IMT}$ may be further reduced.

A weakly metallic CNT (orange circles) may still carry enough current to trigger IMT in the $VO_2$, with $V_{IMT}$ between that of a device with a good metallic CNT (blue) and one without any CNT at all (red).

## 2c. Temperature Dependence

As expected, $V_{IMT}$ reduces as the ambient temperature ($T_0$) increases, which is set by the stage temperature. This is because less power is needed to raise the device temperature to $T_{IMT}$. This is true both for devices with (Figure S6a) and without a CNT. The switching field $E_{IMT}$ (extracted from a linear fit to $V_{IMT}$ vs. $L$) decreases for both types of devices as the temperature increases, as shown in Figure S6b. If the ambient temperature is high enough (above or at $T_{MIT}$), then the device will remain metallic after the bias is removed due to hysteresis in the $VO_2$. At all temperatures, the switching field for devices with a CNT is less than half that required for devices without a CNT.

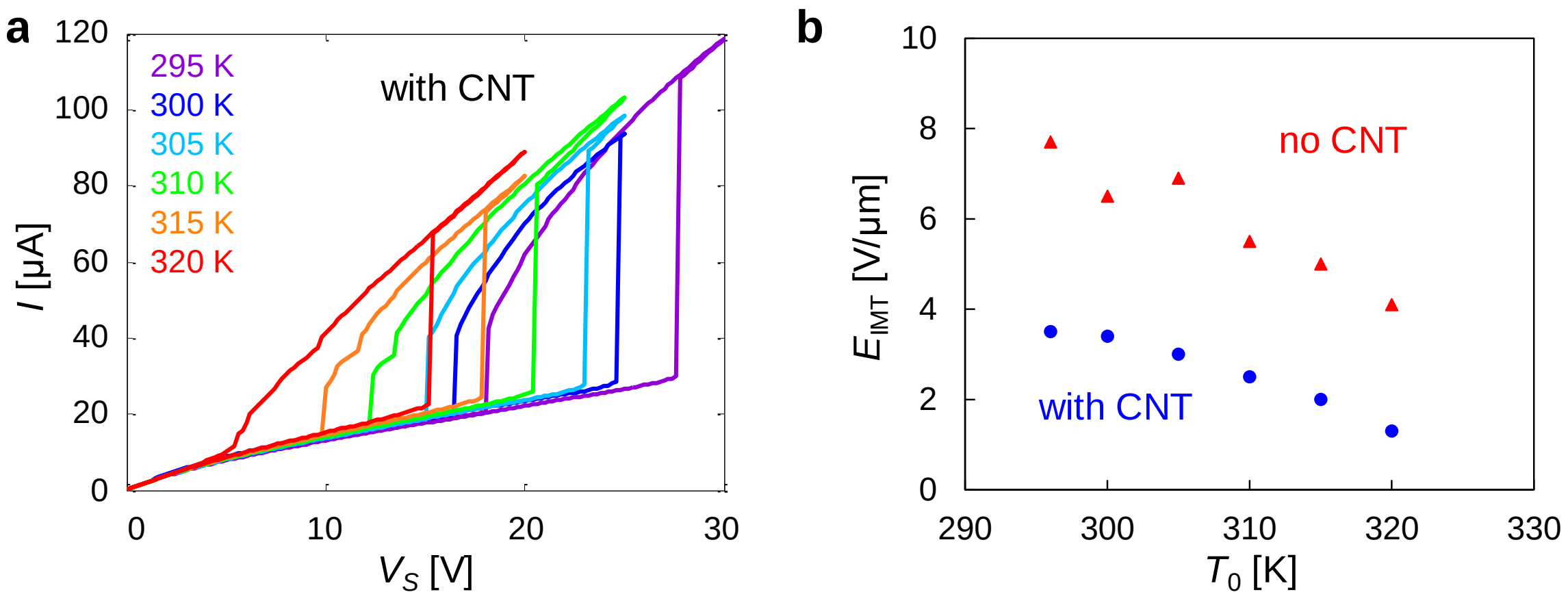

**Figure S6.** (a) Switching of a device with a CNT ($L$ = 5 µm, $W$ = 5.1 µm) using DC voltage control, showing reduced switching voltage as the stage temperature increases. (b) The effective switching field $E_{IMT}$ lowers as temperature is increased, for both devices with (blue circles) and without (red triangles) a CNT. $E_{IMT}$ is extracted as the slope of a linear fit to $V_{IMT}$ vs. $L$ at each stage temperature.

## 2d. Contact Resistance

An estimate of the contact resistance $R_C$ between Pd and insulating $VO_2$ can be obtained using the transfer length method (TLM), shown in Figure S7. The y-intercept corresponds to $2R_C$ and the slope corresponds to the sheet resistance $R_{sh}$. The devices are spread across the chip and have different widths, resulting in significant spread in their resistance since the cleanliness and contact quality can vary, as well as to a lesser extent the $VO_2$ resistivity and thickness. We get a value for $R_{sh}$ of 2520 ± 150 kΩ, and a value for $R_C$ of 460 ± 330 kΩ·µm or equivalently, a contact resistivity $\rho_c \sim (0.8 \pm 0.3) \times 10^{-3}\ \Omega\cdot cm^2$. For a typical 5 µm wide device, this gives a contact resistance of 92 ± 66 kΩ.

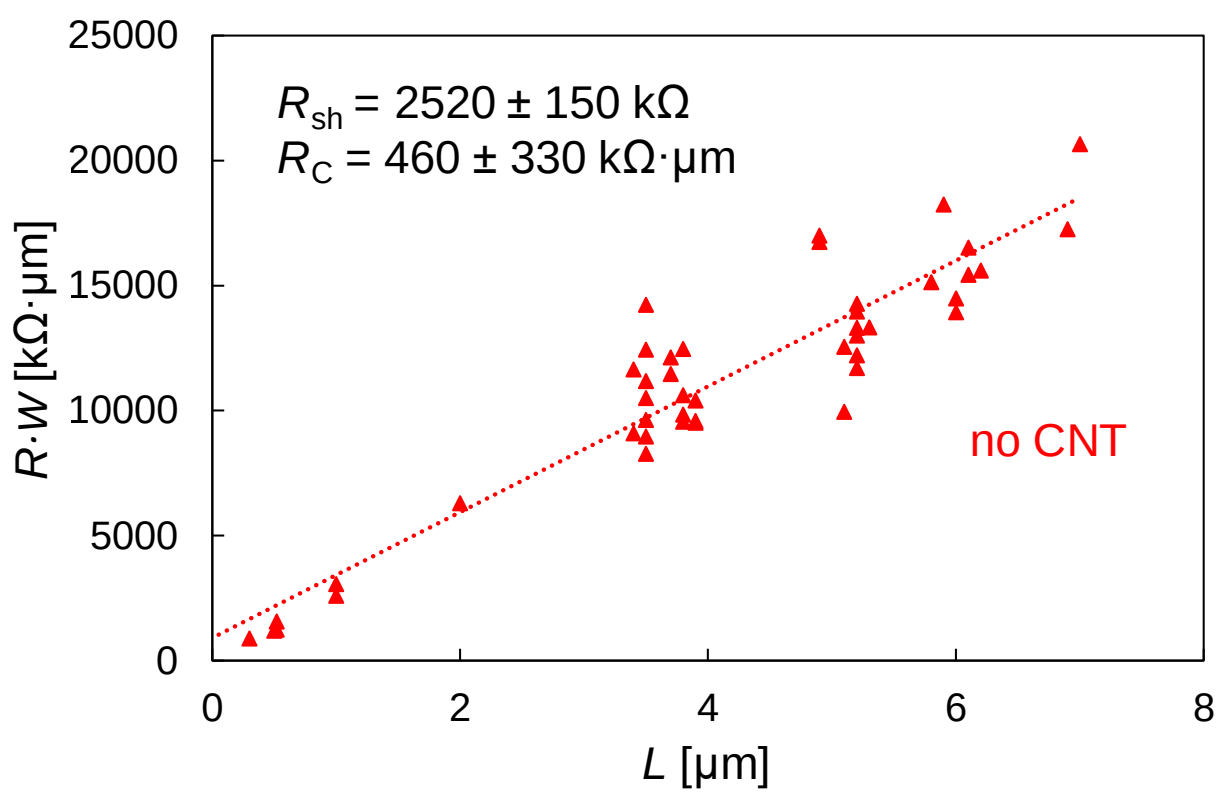

**Figure S7.** Insulating state resistance for $VO_2$ devices without a CNT as a function of length. The vertical axis intercept gives an estimate of twice the device contact resistance ($2R_C$), and the slope gives an estimate for sheet resistance ($R_{sh}$).

## 2e. Series Resistance and Burn-in

A series resistor ($R_S$) is necessary to protect devices from overheating failure in the metallic state (the heat can melt the contacts). The value should be chosen low enough to not affect $V_{IMT}$ (a small fraction of the high-resistance insulating state), but high enough to prevent permanent damage in the metallic state (higher than the low-resistance metallic state). There is no one correct value for $R_S$ and a range of values can be used, as shown in Figure S8a. Using a larger $R_S$ reduces the metallic state current, thus limiting the on/off ratio. There is a slight increase in the total applied voltage ($V_S$) required to switch the device but $V_{IMT}$ does not change, as shown in Figure S8b. As device dimensions are reduced, power dissipation in the metallic state reduces because $V_{IMT}$ decreases. Thus, shorter devices do not need as high a value of $R_S$ for protection.

Figure S8c shows that similar trends are observed for a $VO_2$ device with a CNT heater. However, when higher values of $R_S$ are used, the transition shows an increasingly long period of slow increase in current, prior to a smaller abrupt transition. Akin to the negative differential resistance (NDR) in a current sweep, this slowly rising region corresponds to oscillations. The CNT triggers switching in the $VO_2$, but the subsequent reduction in voltage across the device means there is not enough power needed to remain metallic (the CNT and $VO_2$ cool) so the $VO_2$ returns to insulating, where the CNT heating is again able to switch the $VO_2$, and the system repeatably cycles between IMT and MIT.

Figure S8d shows that as $R_S$ increases, the metallic state resistance ($R_{ON}$) of the device increases (calculated just after switching, with $R_S$ subtracted out). This indicates that increasing $R_S$ reduces the steady state volume of the metallic region or "filament", consistent with prior work.[4] As a secondary effect this change in $R_{ON}$ could be partly due to the metallic state's temperature coefficient of resistance (TCR), since changing $R_S$ changes the device temperature. In devices with a CNT, $R_{ON}$ plateaus when devices show oscillatory behaviour (where $R_{ON}$ is determined just after the final abrupt transition). This may indicate that there is a minimum width (a minimum metallic state power) required for the metallic "filament" to be stable.

Figure S8a shows where we attach the series resistor $R_S$, at the base of the probe tip, to avoid excess cable connections. The location of this resistor is important as well as its value. The abrupt change in device resistance triggers a large transient current due to capacitors discharging that are in parallel with the device (since $I = CdV/dt$).[5,6] This includes intrinsic device capacitance as well as other parasitic capacitances from the contacts and measurement setup such as probe pad capacitance (pad-to-pad or through the $TiO_2$ substrate to the stage) or coax cable capacitance. $R_S$ should be physically close to the device, because then only the device capacitances see the change in conductance (the resistor shields the device from the other external capacitors). A better choice would be to use an integrated on-chip resistor (or transistor, for a variable $R_S$ with smaller area) as a current limiter directly connected to the $VO_2$ device, however our choice in Figure S9a is the best practical one, given the setup.

A typical current overshoot event at the IMT is shown in Figure S9b for a device without a CNT, measured using a 50 Ω oscilloscope in series with the device. Although the overshoot lasts less than 100 ns, the current can transiently reach up to a few mA, which is significant compared to the few hundred μA which is limited by $R_S$ when the device is in the metallic state. The overshoot duration is smaller in shorter devices. Overshoot also occurs on a similar time scale for devices with a CNT, but with a reduced magnitude.

Devices without a CNT show a burn-in effect, displayed in Figure S9c, with the first switching occurring at a slightly higher $V_{IMT}$ than all subsequent switching. This difference is typically on the order of 0.2 – 4 V and is typically higher in narrower devices. This burn-in also results in a slight increase in insulating state conductivity. Burn-in may be a result of the heat from current overshoot at the moment of switching, which could cause a subtle change in the $VO_2$ or the contacts (such as breaking through the native surface oxide or annealing the contacts), making it slightly easier to repeat switching. Devices typically recover from burn-in (if measured again another day then switching voltage has shifted back up for the first switching event, and the device experiences burn-in again). All other switching voltages, currents, and *I*-*V* characteristics reported are after this initial burn-in. Burn-in is not observed in devices with a CNT and is reduced in short devices without a CNT because of the lower switching power and current overshoot.

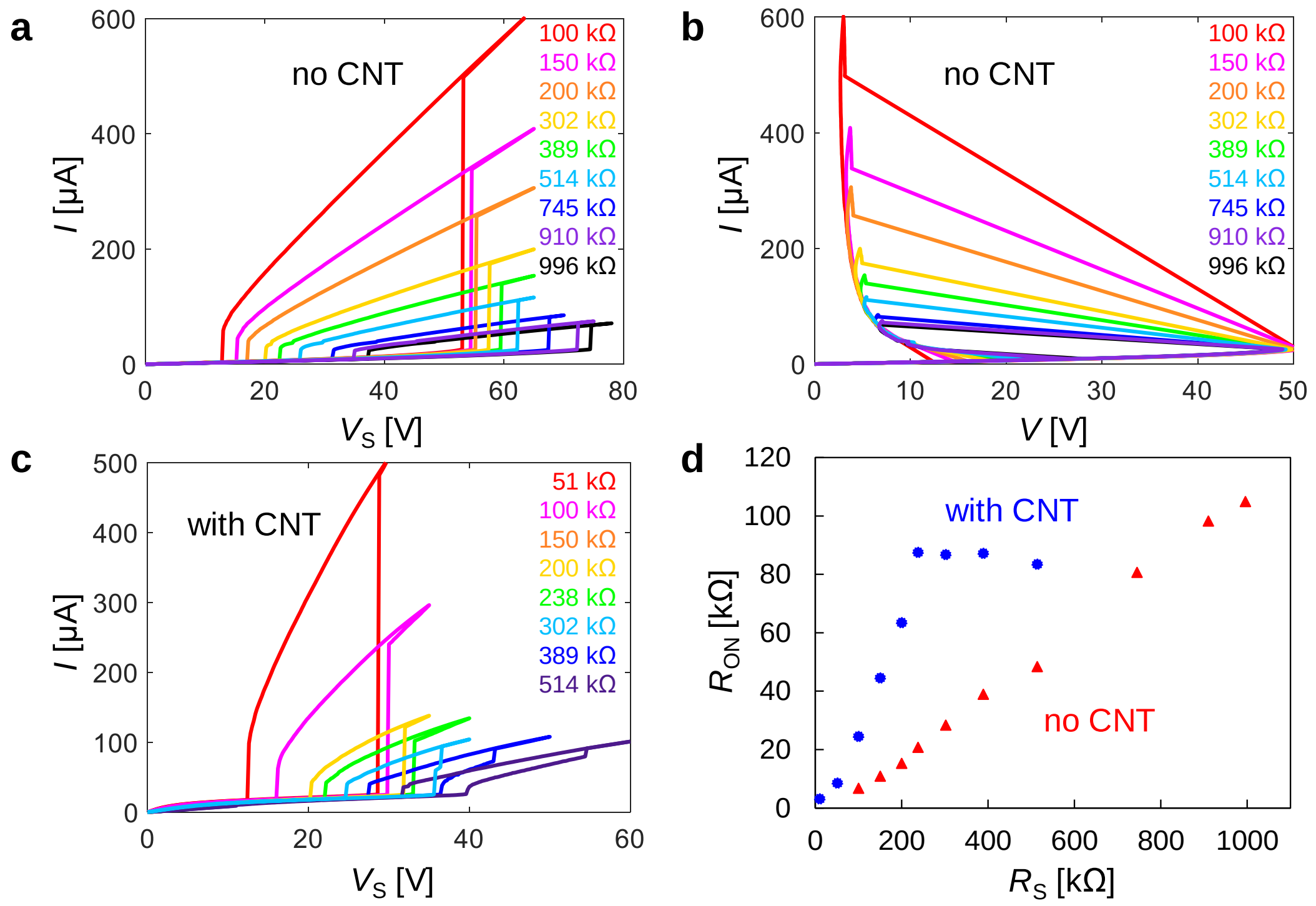

**Figure S8.** (a) DC voltage-controlled switching of a $VO_2$ device without a CNT ($L$ = 5.1 μm, $W$ = 4.1 μm) for different values of $R_S$, plotting total applied voltage $V_S$. (b) The same *I*-*V* measurements as in (a) plotted with the voltage across the device, $V_{IMT}$. The choice of $R_S$ does not impact $V_{IMT}$. (c) DC voltage-controlled switching of a device with a CNT ($L$ = 7.1 μm, $W$ = 2.8 μm) for different $R_S$. (d) Device resistance just after switching ($R_{ON}$), as a function of $R_S$ for devices without (red triangles) and with a CNT (blue circles). $R_S$ can be used to control the metallic state "filament" width.

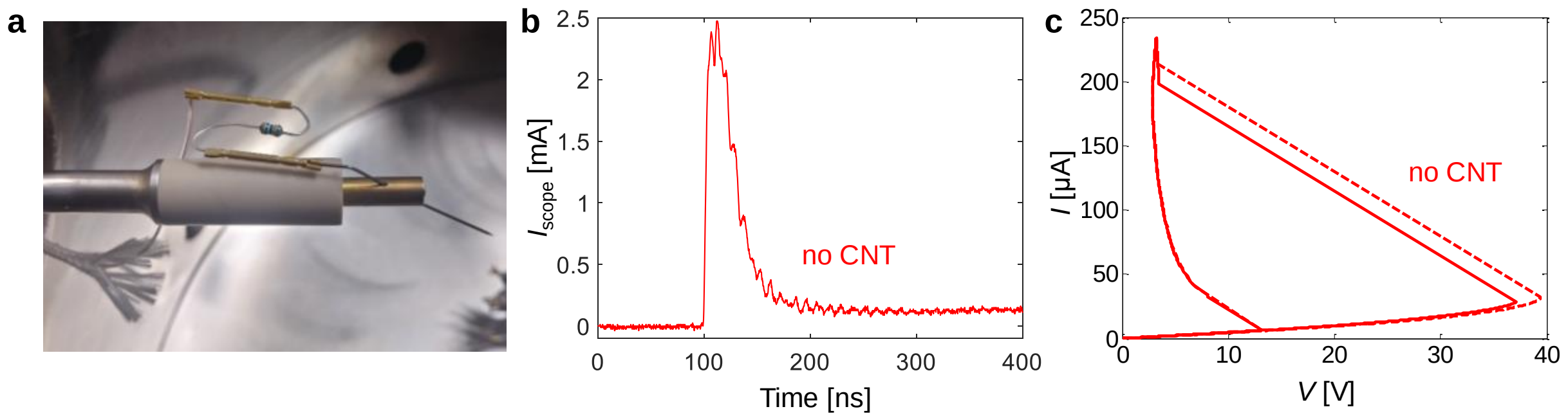

**Figure S9.** (a) Setup for using a series resistor $R_S$ as current compliance, with $R_S$ on the probe tip, as close as possible to the device under test. (b) Typical current overshoot when a device without a CNT switches ($L$ = 3.5 μm, $W$ = 2.7 μm, due to capacitive discharge. The current is measured using a 50 Ω oscilloscope in series with the device. Devices with a CNT also show current overshoot waveforms with a nearly identical shape and time-scale, but with a reduced magnitude. (c) DC voltage-controlled switching of a device without a CNT ($L$ = 3.8 μm, $W$ = 5.1 μm) on the first sweep (dashed), and subsequent sweeps (solid). There is a slight shift (a "burn-in") in $V_{IMT}$ after the first switching event.

### 2f. Switching Field

Thermally triggered switching occurs when the steady state temperature in the device (or some part of it) reaches the insulator-metal transition temperature ($T_{IMT}$). According to a very simple Joule heating description of a $VO_2$ block on a substrate, the input electrical power $P = V^2/R$ causes the temperature to rise with a maximum determined by the thermal conductance to the surrounding environment $G_{th}$:

$$\frac{V^2}{R} = G_{th}(T_{IMT} - T_0) \tag{1}$$

If we write an effective switching field $E_{IMT} = V/L$, where $V$ is the voltage across the $VO_2$ length $L$ at the transition (excluding the voltage drop at the contacts), then the field could be described as:[6]

$$E_{IMT} = \sqrt{\frac{G_{th}R(T_{IMT} - T_0)}{L^2}} \tag{2}$$

This does not capture aspects of real devices like the contacts and non-uniform heating. However, it is useful in pointing out some dependencies of the switching field. Via $G_{th}$ the switching field depends on the thermal boundary resistance to the substrate, the thermal conductivity of the substrate, the geometry, and the resistivity of the $VO_2$. Two-terminal devices can vary in this regard, making it hard to give a direct comparison of switching field. Table 1 lists switching fields of two-terminal switching devices extracted from literature with a similar change in temperature to our work ($\Delta T = T_{IMT} - T_0 \approx 32$ K). Fields are listed as a slope (with a vertical axis intercept $V_{offset}$ due to heat loss and resistance at the contacts) or as a long channel $V/L$ where $L$ is at least a few microns long. We note that thicker, wider $VO_2$ devices (lower $R$) on better thermal insulating substrates (lower $G_{th}$) typically have lower switching field $E_{IMT}$.

Our switching fields and intercepts fall within these ranges, although our field is on the higher end (see Figure 2c and Figure S6b, i.e. between ~3.5 and 7 V/μm for devices with and without a CNT, around $T_0$ = 296 K). This is likely because our $VO_2$ films are single crystals and quite thin (higher $R$) compared to others, and because our $VO_2$ has better thermal conductance (higher $G_{th}$) to the crystalline $TiO_2$ substrate.

**Table 1.** Switching Fields in Prior $VO_2$ Films

| Ref. | Deposition Technique | Substrate | $L$ (µm) | $W$ (µm) | $t$ (nm) | Contact | $E_{IMT}$ (V/µm) | $V_{offset}$ (V) | $\Delta T$ (K) |
|---|---|---|---|---|---|---|---|---|---|
| 7 | PLD | $TiO_2$ (001) | 0.08 – 0.11 | 1 | 10 | Ta/Au | 13 | 0.1 | 35 |
| 8 | MBE | $TiO_2$ (001) | 6 | 10 | 10 | Pd | 2.2 | - | 35 |
| 9 | PLD | $SiO_2$ (500 nm) | 2, 10 | 20 | 100 | Cr/Cu/Au | 1.2 – 4.3[a] | - | 51 |
| 10 | sputtered | $SiO_2$ (2 µm) | 0.1 – 1 | 10 | 100 | Ag, Cu, Au, Pd | 2.0 – 2.3[b] | 2.2 – 3.3[b] | 45 |
| 11 | sputtered | $SiO_2$ (500 nm) | 7.5 | 50 | 200 | Al | 0.5 | - | 34 |
| 12 | sputtered | $SiO_2$ (100 nm) | 4 | 4 | 100 | Pt/Au | 1.2 | - | 28 |
| 13 | oxidation | poly-$Al_2O_3$ (20 nm) | 0.1 – 5 | 2, 10 | 100 | Ti/Au | 17.4 | 4.1 | 53 |
| 4 | oxidation | $Al_2O_3$ $(11\bar{2}0)$ | 5 – 80 | 100 | 200 | Cr/Au | 2.9<br>2.2 | 24.4<br>22.0 | 37<br>27 |
| 14 | PLD | $Al_2O_3$ (1102) | 5 | 50 | 90 | Cr/Au | 2.3, 3.4[c] | - | 32 |
| 15 | sputtered | $Al_2O_3$ $(10\bar{1}0)$ | 20 | 100 | 100 | Ni | 1.0 | - | 45 |
| 16 | sputtered | $Al_2O_3$ (r-cut) | 20 | 50 | 130 | Au | 2.6 | - | 18 |
| 17 | sol-gel | $Al_2O_3$ | 10 | 10 | 100 | Ni | 1.0 | - | 45 |

[a]different contact geometries, [b]different contact metals, [c]different voltage polarity

## 3. Kelvin Probe Microscopy (KPM)

During each KPM scan the device is held at a constant voltage. The current remains constant at each bias point and is shown for the $VO_2$ device without a CNT in Figure S10a and with a CNT in Figure S10b. Because $R_S$ = 200 kΩ is used, upon IMT the device voltage snaps back to ~5 V and ~8 V for devices with and without a CNT, respectively. The magnitude of the linear potential drop observed across the channel at each bias prior to flattening is shown for a device without a CNT in Figure S10c and with a CNT in Figure S10d. There is a slightly negative offset in the potential, which is determined by the workfunction difference between the tip and sample, as well as any error in the digital-to-analog converter.

Figures S11 and S12 show the full set of KPM scans taken with increasing $V_S$ for the devices without and with a CNT, respectively, with labels the same as in Figure S10. These images are obtained by using a first order line flattening operation. Although this CNT had a small kink, there is no additional potential drop across it, so it is not significantly more resistive and therefore not the main cause of the heating and IMT in the $VO_2$.

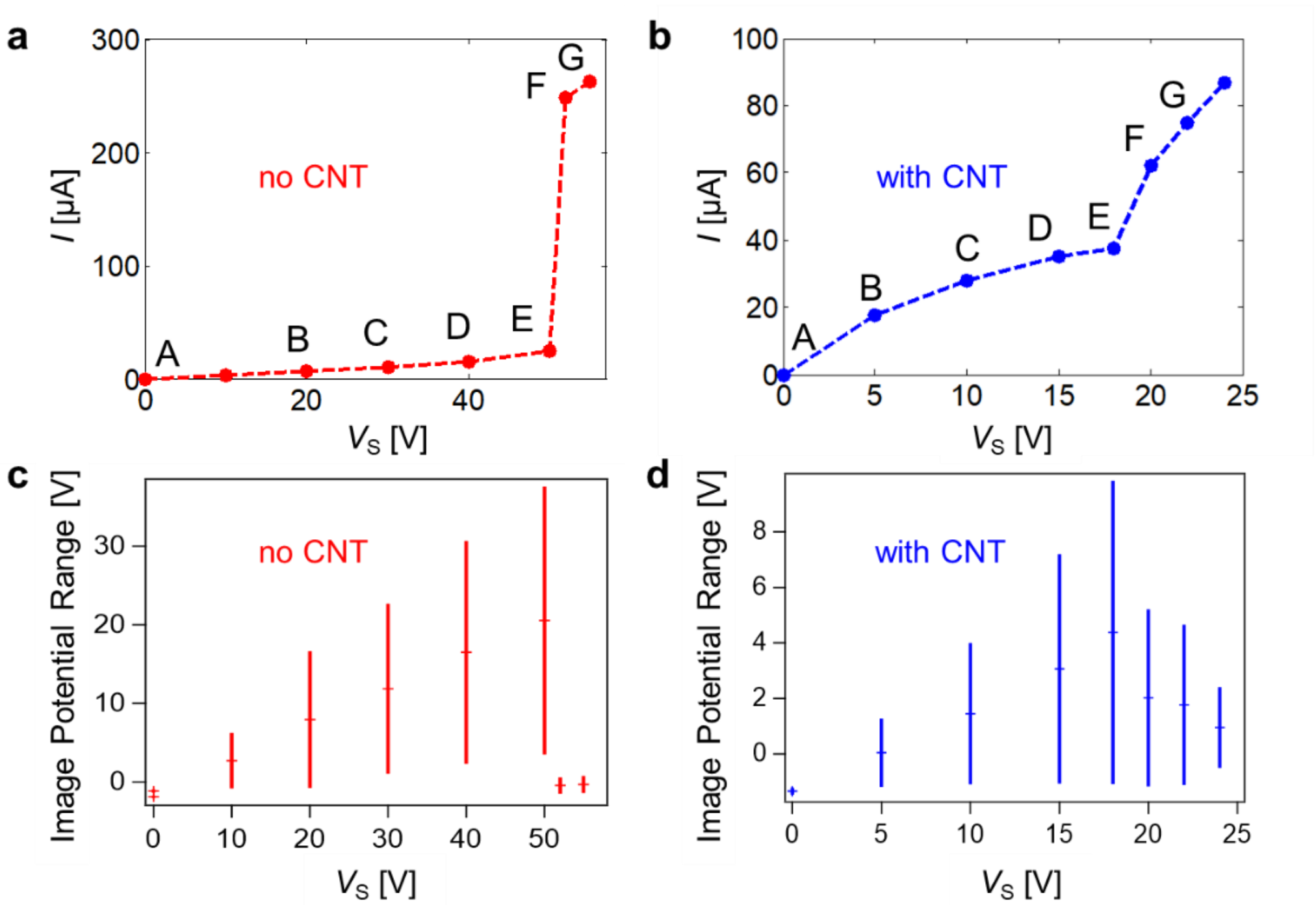

**Figure S10**. Current at each applied voltage $V_S$ during KPM measurements of devices without (a) and with a CNT (b). Raw potential drops measured across devices without (c) and with a CNT (d) prior to flattening. The potential drops after switching because of the device voltage snapback caused by $R_S$.

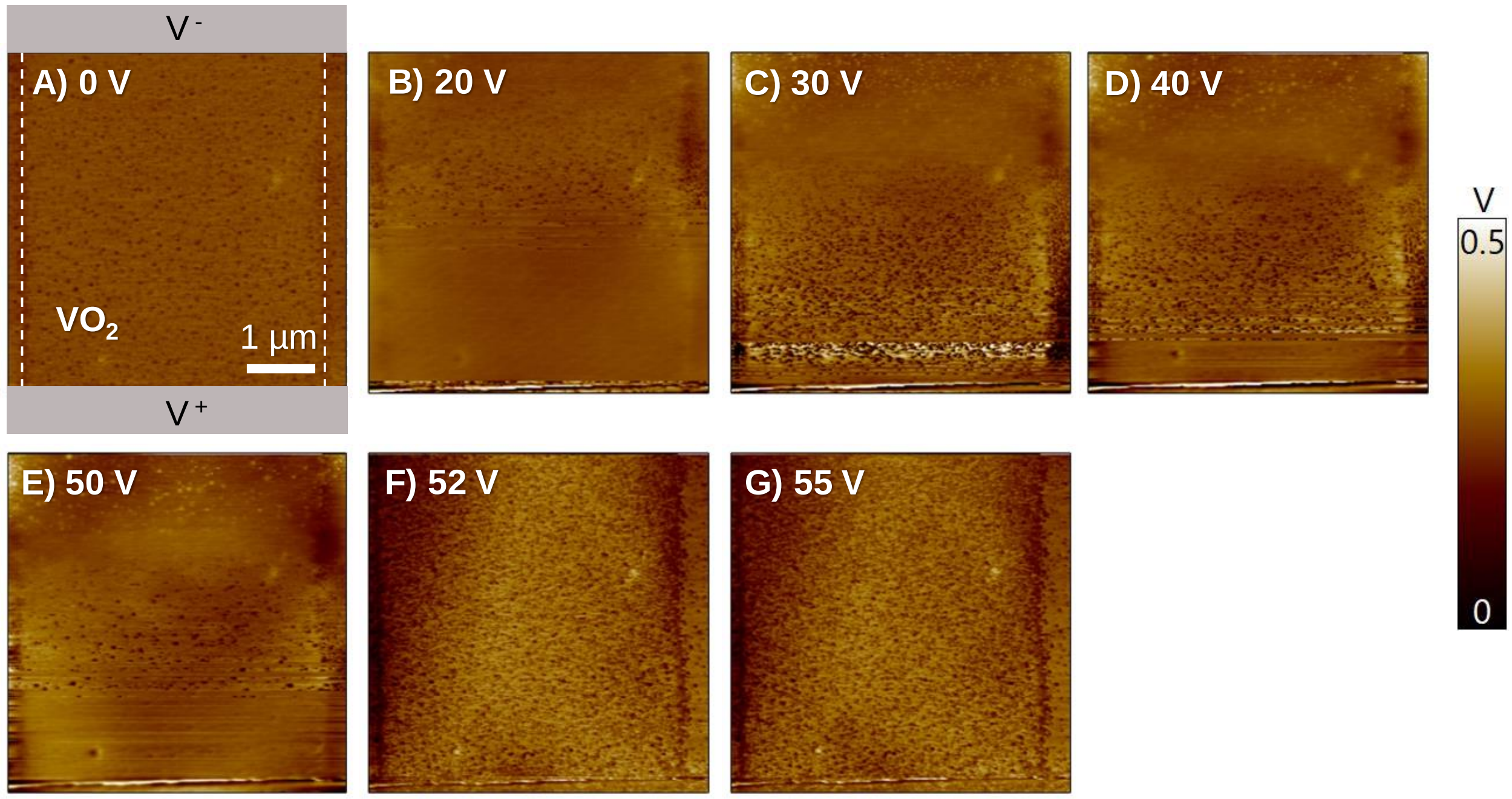

**Figure S11.** Full set of KPM images of a device without a CNT. See Figure S10a for corresponding *I*-*V* data. (A) is with no bias, (B) – (E) are in the insulating state, and (F) - (G) correspond to the metallic state, with applied $V_S$ as labeled. The electrodes are just outside the images, at the top and bottom, as labelled in (A). The edges of the $VO_2$ channel are marked in (A) with dashed lines.

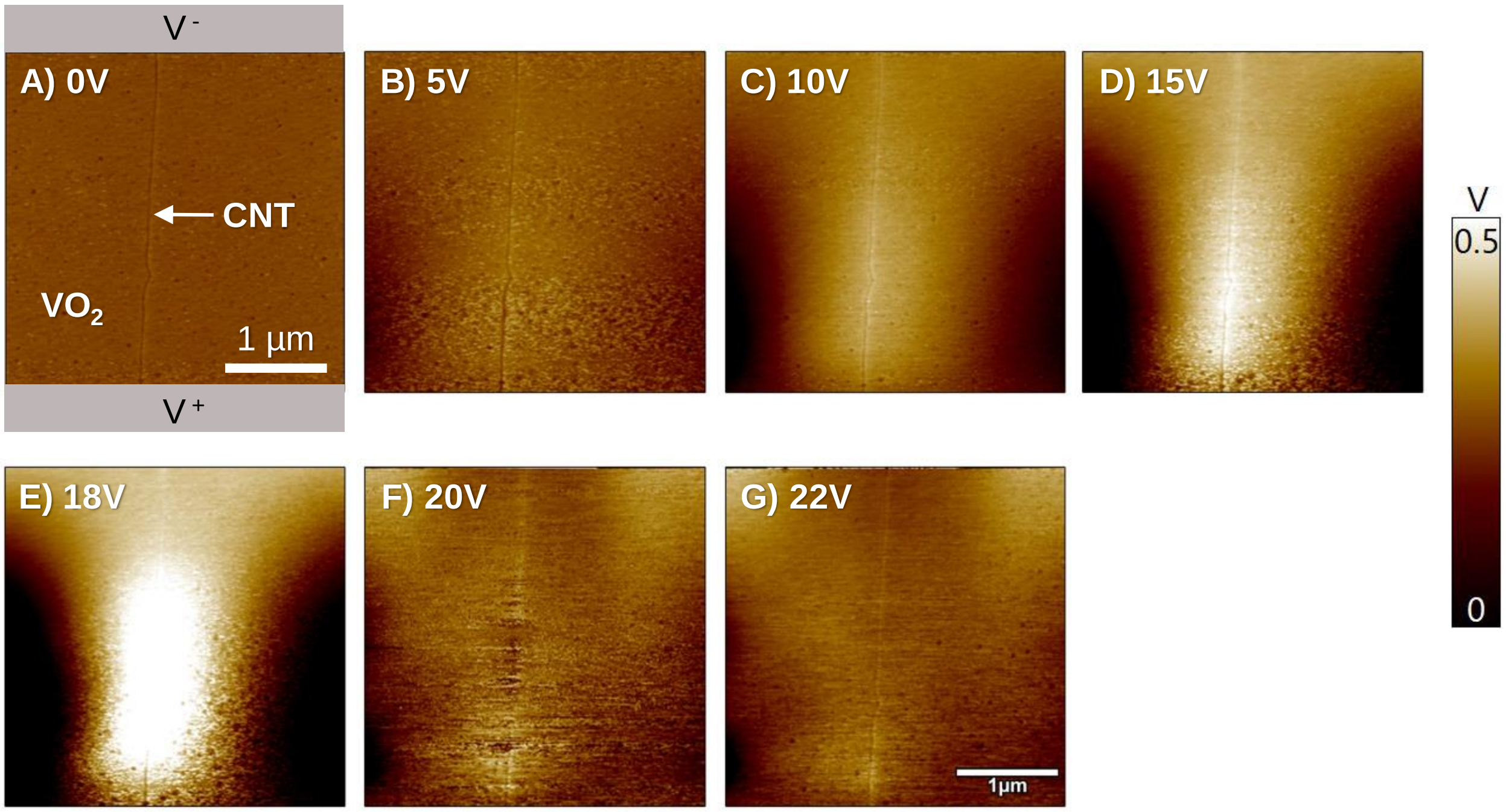

**Figure S12.** Full set of KPM images of a device with a CNT. See Figure S10b for corresponding *I*-*V* data. (A) is with no bias, (B) - (E) are in the insulating state, and (F) - (G) correspond to the metallic state, with applied $V_S$ as labeled. The electrodes are just outside the images, at the top and bottom, as labelled in (A). The $VO_2$ channel edges are just outside the images, on the left and right.

## 4. Three-Dimensional Finite Element Simulations

### 4a. Model Details

We developed a three-dimensional (3D) finite element model for our devices using COMSOL Multiphysics, which self-consistently considers electrical and thermal effects. An electrical model is used to calculate the voltage and current distribution in the device, while a thermal model determines the temperature distribution. These are coupled together via Joule heating, as well as by the temperature dependent resistivities of the CNT and $VO_2$.

The model solves the following system of equations, where Eqn. 1 is used to obtain the voltage and current distribution in the device while Fourier's law of heat conduction in steady state (Eqn. 2) is used to obtain the local temperatures. $k$ is the thermal conductivity, and σ is the electrical conductivity which is temperature dependent (and therefore spatially dependent) for the $VO_2$ and CNT. These two equations are coupled together via $J \cdot E$ as a Joule heating source term, with $J$ being the local current density and $E$ the local electric field.

$$\nabla \cdot (\sigma(x,y,z,T)\nabla V) = 0 \qquad (1)$$

$$\nabla \cdot (k\nabla T) + J \cdot E = 0 \qquad (2)$$

For the $VO_2$, $\sigma(T)$ is based on measurements of $R(T)$ from our films as a function of stage temperature (main text Figure 1i). The $R(T)$ data is imported as a table of temperature and resistance with ~1 K steps. When solving it is important to use nearest-neighbour as the interpolation type, which will round the temperature of an element to the nearest table value and use the corresponding resistance. If the default linear interpolation is used then a slight change in temperature during the IMT (the steepest part of $R(T)$) while solving will cause a large change in resistance between solver steps, greatly hindering convergence. Only one branch (heating or cooling) of the $R(T)$ curve is used at a time for each voltage point simulated. For generating an *I-V* curve with hysteresis included, the heating branch was used as voltage was increased, and the cooling branch was used after switching as the voltage was swept back down. The solution for the previous voltage point was used as the initial conditions for the next voltage.

The conductivity of the CNT is based on a model developed by Pop *et al*,[3] given by:

$$\sigma_{\mathrm{CNT}}(T, V, L) = \frac{4q^2}{h}\frac{\lambda_{\mathrm{eff}}}{A} \tag{3}$$

$$\text{where } \lambda_{\mathrm{eff}} = \left(\lambda_{\mathrm{AC}}^{-1} + \lambda_{\mathrm{OP,ems}}^{-1} + \lambda_{\mathrm{OP,abs}}^{-1} + \lambda_{\mathrm{defect}}\right)^{-1} \tag{4}$$

$\lambda_{\mathrm{eff}}$ is an effective electron mean free path (MFP) obtained using Matthiessen's rule. It combines contributions from elastic electron scattering with acoustic phonons ($\lambda_{\mathrm{AC}}$), and inelastic electron scattering by optical phonon absorption ($\lambda_{\mathrm{OP,abs}}$) and emission ($\lambda_{\mathrm{OP,ems}}$). Emission is influenced by the electric field, so this term is dependent on the applied voltage and CNT length, and all MFPs are a function of temperature. Values of $\lambda_{\mathrm{OP,300}}$ = 20 nm and $\hbar w_{op}$ = 0.2 eV are used in the model, which is described in detail in Ref. 3. An additional scattering term for defects ($\lambda_{\mathrm{defect}}$), with a mean free path of 0.8 µm, is added to better represent *I-V* characteristics of our imperfect CNTs. The reduction in CNT conductivity with increasing temperature results in current saturation behaviour of the CNT, described in Ref. 3 and seen in Figure S5a.

Figure S13a shows the geometry for the simulation corresponding to the device with a CNT used for SThM in main text Figure 4, and Figure S13b shows the geometry for the device simulations in main text Figure 5. The CNT is centered in the device, so only half the device is simulated due to symmetry. The CNT is approximated as a rectangular prism 1.2 nm wide and 1.2 nm tall on top of the 5 nm thick $VO_2$, spanning the entire length of the device and underneath the 50 nm thick Pd contacts. The simulated $TiO_2$ substrate is 2 µm thick (unlike the ~500 µm experimental $TiO_2$), which is sufficient to capture its thermal resistance, most of which occurs at the thermal constriction near the CNT. For $VO_2$ devices without a CNT, where there is significantly larger volume being heated, a 15 µm thick $TiO_2$ substrate needs to be simulated to accurately determine the device temperature. The device for SThM is 6 µm wide and 7 µm long, with the $VO_2$ covered by a 50 nm thick layer of poly(methyl methacrylate) (PMMA). The other simulated $VO_2$ devices with and without a CNT are 4 µm wide and 5 µm long, with no PMMA capping. For these devices, a 200 kΩ series resistor $R_S$ is modeled as a 50 nm thick resistive layer on top of one of the contacts, with a resistivity chosen to give a total resistance of 200 kΩ. The $R_S$ layer does not thermally interact with the device. Including the extra width of $TiO_2$ beyond the $VO_2$ width makes no significant difference to devices with a CNT (whose $VO_2$ edges are near room temperature far from the CNT heating), but a large width of $TiO_2$ is needed to fully capture the thermal profile in devices without a CNT. Material parameters are listed in Table 2.

The top of one contact is grounded, and the top of the other (or the top of $R_S$, if applicable) is set at a constant potential. Electrical contact resistance is simulated on internal boundaries between CNT/Pd (25 kΩ), CNT/$VO_2$ (100 kΩ) and $VO_2$/Pd. The $VO_2$/Pd contact resistivity is set to $0.8\times10^{-3}$ Ω·cm$^2$ (see Section 2d) at room temperature and scaled with temperature in the same way as the $VO_2$ resistivity (by using the normalized heating branch of $R(T)$ for the forward voltage sweep and the normalized cooling branch for the backward voltage sweep).[18] Other boundaries are modeled as electrically insulating.

The bottom of the $TiO_2$ is fixed at room temperature (296 K). The top of the device is assumed to be thermally insulating (adiabatic), and the sides of the device have open boundary conditions. The Pd resistivity is taken from measurements of our films, and its thermal conductivity is estimated from the Wiedemann-Franz law. Thermal boundary resistance is simulated on all interior boundaries. These interfacial resistances cannot be ignored, since they limit the flow of heat (especially from the CNT to the $VO_2$ but also from the $VO_2$ to the $TiO_2$ substrate), impacting the local $VO_2$ temperature and therefore the switching voltage. The CNT/$VO_2$ interface is set to $5\times10^{-9}$ $m^2K/W$, the $TiO_2/VO_2$ interface to $8\times10^{-9}$ $m^2K/W$, and all other interfaces to $10^{-8}$ $m^2K/W$. To the authors' knowledge these have not been experimentally measured, so were adjusted within reason[19] to better fit the *I-V* characteristics of the $VO_2$ device with a CNT used for SThM in main text Figure 4e.

All simulations were solved with a segregated approach, which solves each physics (temperature and voltage) sequentially, repeating until convergence is achieved. The matrix equation containing the description of the physics (Eqn. 1 or 2) on each mesh element is solved using a direct approach (as opposed to iterative) with the MUMPS solver. The temperature was solved as the first step, followed by the electric potential. To help with convergence when solving across the IMT, the electric potential step was given two iterations instead of the default one (the solver will solve temperature, electric field, electric field again, temperature, etc. repeating until convergence). To also help simulate the IMT, the damping factors were set to 0.1 and 0.4 for the temperature and potential segregated steps respectively. The damping factor is a scaling factor reducing the size of the step taken at each iteration of the physics (rather than using the current step's solution as the next starting point, which would correspond to a damping factor of 1). If these factors are not reduced, then the solver will overshoot the solution and oscillate between insulating and completely metallic (the entire device width switched, with too much heat and current), since the problem is nonlinear.

**Table 2.** Material properties used in simulation

| | **σ [S/m]** | **k [$Wm^{-1}K^{-1}$]** |
|---|---|---|
| **$TiO_2$** | $10^{-7}$ | 8 |
| **$VO_2$** | Fig. 1i, function of $T(x,y,z)$ spanning 80 to $2\times10^6$ | 5 |
| **CNT** | Eqns. 3,4 | 600 |
| **Pd** | $2.9\times10^6$ | 23 |
| **PMMA** | $10^{-10}$ | 0.2 |

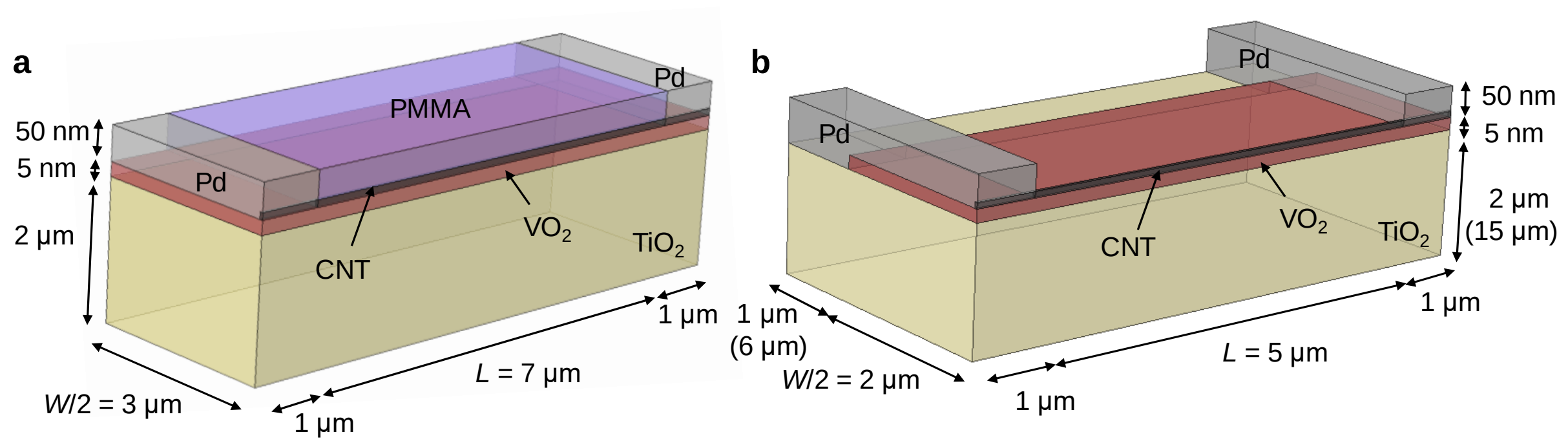

**Figure S13.** (a) Half-device structure used to simulate the experimental device used for SThM in main Figure 4. (b) Half-device structure used to simulate switching and full *I-V* curves in main Figure 5. For the device without a CNT a larger $TiO_2$ substrate is used, with dimensions indicated in brackets.

Simulated temperatures on the surface of a $VO_2$ device without a CNT are shown just before (Figure S14a) and just after (Figure S14b) IMT, corresponding to the simulated *I*-*V* curve in main text Figure 5c. Before switching there is significant bulk Joule heating, and the device requires much more power for switching than with a CNT. A ~400 nm wide metallic "filament" is formed upon IMT in the center of the $VO_2$ device, much wider than the ~10 nm wide "filament" in the device with a CNT. In practical devices, the filament is likely larger due to transient current overshoot, caused by parasitic capacitors discharging at the moment of switching, combined with hysteresis.

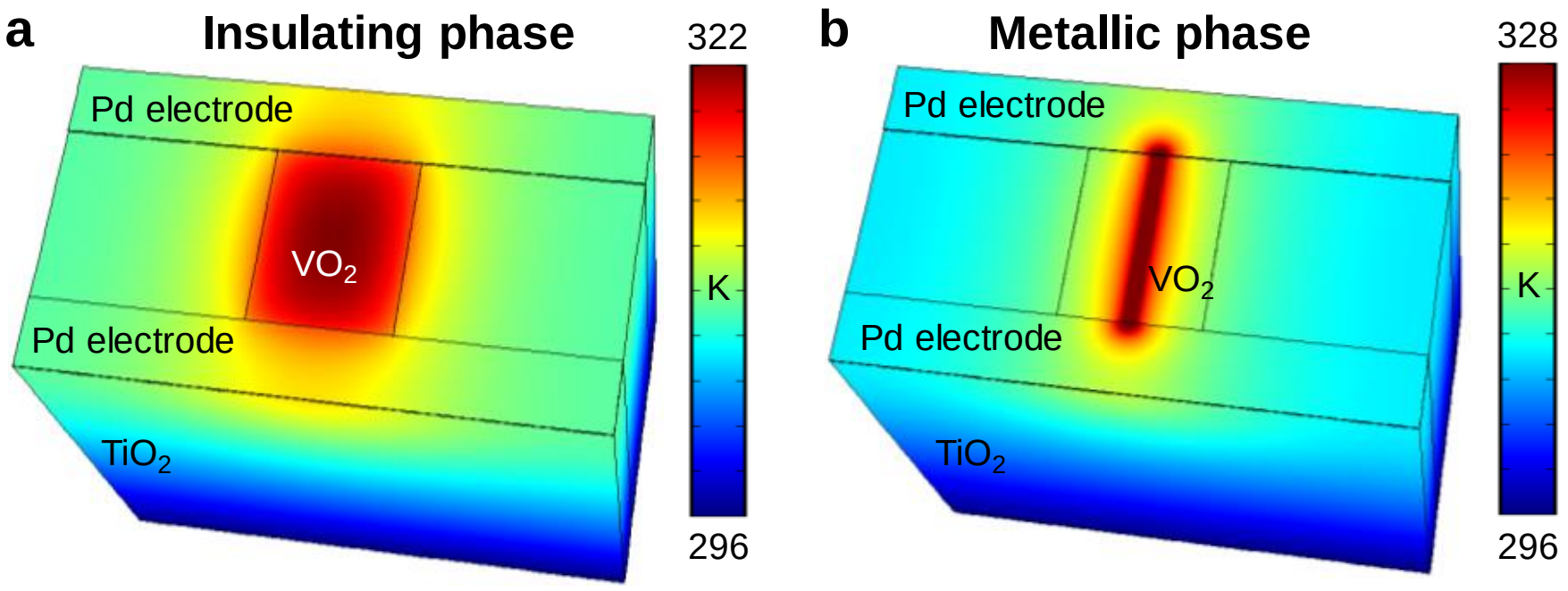

**Figure S14.** Simulated temperature on the surface of a $VO_2$ device without a CNT ($L$ = 5 μm, $W$ = 4 μm) just before (a) and just after (b) the metallic transition, with $R_S$ = 200 kΩ. $VO_2$ requires a higher switching power than devices with a CNT because it relies on bulk Joule heating. After switching, a metallic region forms in the center of the device, where it is hottest, with its size determined by current overshoot and $R_S$.

## 4b. Simulated Device Scaling

Simulated switching voltage $V_{IMT}$, current $I_{IMT}$, and power $P_{IMT}$ of $VO_2$ devices without a CNT are shown in Figure S15 for a variety of device lengths and widths. As device length is reduced (Figure S15a-b), both $P_{IMT}$ and $V_{IMT}$ decrease nearly linearly. $I_{IMT}$ has little length dependence, with only a slight increase in very short devices. These scaling trends match well with what is seen experimentally (Figure 2c-d and Figure S3-4) and reaffirm the thermal dependence of electrical switching in $VO_2$.

As device width is reduced (Figure S15c-d), contact resistance increases and $V_{IMT}$ increases steeply below ~ 1 μm. However, $I_{IMT}$ also decreases, resulting in a net decrease in $P_{IMT}$ for narrower devices. In contrast, $VO_2$ devices with a CNT have no width dependence on switching characteristics. Given these trends, adding a CNT to an ultra-narrow $VO_2$ device would have the benefit of maintaining a much lower, width-independent switching voltage (see Table 3 for a simulated example). However, because the CNT is more conductive than an ultra-narrow stripe of $VO_2$, this comes at the cost of increased insulating state current and therefore reduced on/off ratio. This could result in little improvement in overall switching power, if any. To retain access to a higher resistance state with low current, a gateable semiconducting CNT or other nonlinear heat source (e.g. nanowire, diode, switching material, etc.) could be used instead.

In thermally triggered devices, there is a tradeoff between low switching voltage and current because enough total power is still required to reach the transition temperature. The addition of an extra Joule heating source with an appropriate current density can shift device operation to a lower switching voltage instead of low current. Some additional control over switching voltage could be achieved by engineering the thermal environment of the device, for example, by changing the thermal boundary resistance to and the thermal conductivity of the substrate to reduce heat loss from the $VO_2$. If this improves heating efficiency (*e.g.* by confining heating to a small volume), then a lower switching power can be achieved.

These simulated scaling trends are estimates, assuming similar electrical and thermal transport at small length scales. In reality, there may be some differences when devices are shrunk to the nanoscale. For example, thermal and electrical conductivity of the $VO_2$ will likely decrease when narrower (due to surface boundary scattering), which could affect the switching trends.

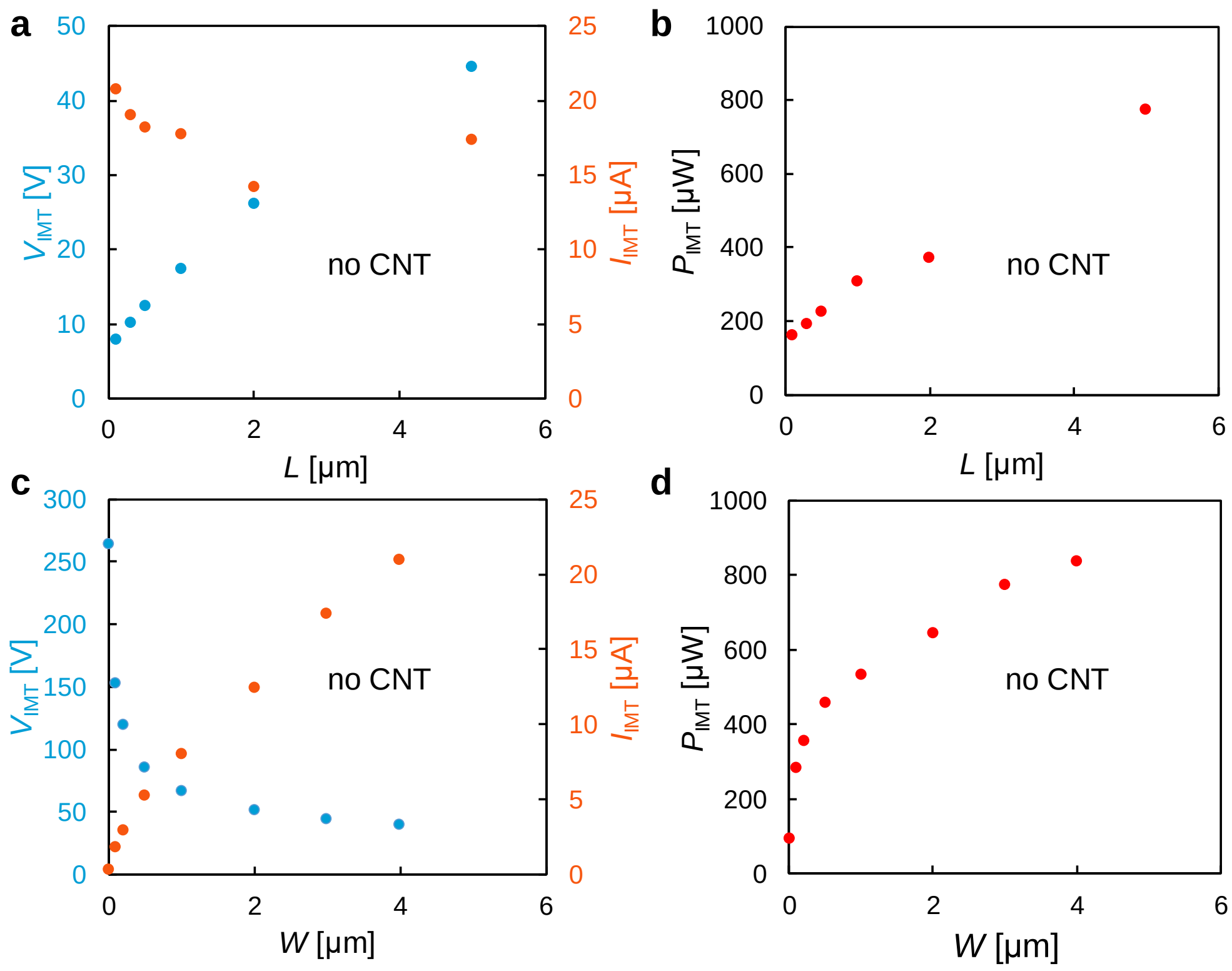

**Figure S15.** Simulated scaling trends of $VO_2$ devices without a CNT, with thermally induced switching. (a) Switching voltage $V_{IMT}$ decreases with device length (simulated down to 100 nm length, with $W$ = 5 μm) and switching current $I_{IMT}$ starts to increase slightly below ~ 1 μm. (b) Total device power $P_{IMT}$ reduces with linearly with length. (c) In narrow devices (simulated down to 10 nm width), $V_{IMT}$ is much higher but $I_{IMT}$ decreases ($L$ = 5 μm). (d) Switching powers decreases as device width is reduced ($L$ = 5 μm).

**Table 3.** Simulated Switching of Narrow Devices ($L$ = 5 μm, W = 100 nm)

| | **without CNT** | **with CNT** |
|---|---|---|
| **$V_{IMT}$ [V]** | 153 | 9 |
| **$I_{IMT}$ [μA]** | 2 | 33 |
| **$P_{IMT}$ [μW]** | 282 | 528 |

**Table 3.** Simulated switching of long, narrow $VO_2$ devices with and without a CNT. The CNT greatly reduces the switching voltage but comes with a higher switching current and power (although the simulated CNT is more conductive than imperfect experimental CNTs). By also reducing the device length, the switching voltage and power would decrease for both types of devices.

## 5. Supplementary Video

Video S1 (uploaded separately) shows the evolution of surface temperature of the simulated device with a CNT (main Figure 5a), as the voltage is swept up and then down on the *I-V* curve (main Figure 5b).

## 6. Supplementary References